\documentclass[twocolumn,times, twocolappendix]{aastex7}

\definecolor{malachite}{rgb}{0.04, 0.85, 0.32}

\usepackage{soul,xcolor}
\setstcolor{red}

\usepackage{subfigure}
\usepackage{amsmath}
\usepackage{chngcntr}
\begin{document}

\title{Spatially Resolved, Multiphase Mass Outflows of the Seyfert 1 Galaxy NGC 3227}

\author[0000-0001-7238-7062]{Julia Falcone}
\affiliation{Department of Physics and Astronomy, Georgia State University, 25 Park Place, Atlanta, GA 30303, USA}
\email{jfalcone2@gsu.edu}

\author[0000-0002-6465-3639]{D. Michael Crenshaw}
\affiliation{Department of Physics and Astronomy, Georgia State University, 25 Park Place, Atlanta, GA 30303, USA}
\email{dcrenshaw@gsu.edu}

\author[0000-0002-4917-7873]{Mitchell Revalski}
\affiliation{Space Telescope Science Institute, 3700 San Martin Drive, Baltimore, MD 21218, USA}
\email{mrevalski@stsci.edu}

\author[0000-0002-3365-8875]{Travis C. Fischer}
\affiliation{AURA for ESA, Space Telescope Science Institute, 3700 San Martin Drive, Baltimore, MD 21218, USA}
\email{tfischer@stsci.edu}

\author[0000-0001-8658-2723]{Beena Meena}
\affiliation{Space Telescope Science Institute, 3700 San Martin Drive, Baltimore, MD 21218, USA}
\email{bmeena@stsci.edu}

\author[0009-0005-3001-9989]{Maura Kathleen Shea}
\affiliation{Department of Physics and Astronomy, Georgia State University, 25 Park Place, Atlanta, GA 30303, USA}
\email{mshea3@gsu.edu}


\author[0000-0002-2713-8857]{Jacob Tutterow}
\affiliation{Department of Physics and Astronomy, Georgia State University, 25 Park Place, Atlanta, GA 30303, USA}
\email{jtutterow1@gsu.edu}

\author[0000-0003-3401-3590]{Zo Chapman}
\affiliation{College of Computer Science, Georgia Institute of Technology, 266 Ferst Drive, Atlanta, GA 30332, USA}
\email{zoechapman147@gmail.com}


\author[0009-0005-2145-4647]{Kesha Patel}
\affiliation{Department of Physics and Astronomy, Emory University, 400 Dowman Drive,
Atlanta, GA 30322, USA}
\email{kesha.patel@emory.edu}

\correspondingauthor{Julia Falcone}
\email{jfalcone2@gsu.edu}


    
\begin{abstract}
We present spatially resolved mass outflow rates of the ionized and molecular gas in the narrow line region of the Seyfert 1 galaxy NGC 3227. Using long-slit spectroscopy and [O~III] imaging from from Hubble Space Telescope's Space Telescope Imaging Spectrograph and Apache Point Observatory’s Kitt Peak Ohio State Multi-Object Spectrograph, in conjunction with Cloudy photoionization models and emission line diagnostics, we find a peak ionized mass outflow rate of $\dot M_{\text{ion}} =$ $19.9\pm9.2$ M$_\odot$ yr$^{-1}$ at a distance of $47\pm6$ pc from the supermassive black hole (SMBH). Using archival data from the Gemini-North Near-infrared Field Spectrograph measuring H$_2$~$\lambda2.1218$ $\mu$m emission, we find a maximum peak warm molecular outflow rate of $\dot M_{\mathrm{H_2}} \le 9 \times 10^{-4}$ M$_\odot$ yr$^{-1}$ at a distance of $36\pm6$ pc from the SMBH. Using archival data from the Atacama Large Millimeter/submillimeter Array measuring CO(2-1) emission, we find a maximum peak cold molecular gas mass outflow rate of $\dot M_{\mathrm{CO}} \le$ $23.1$ M$_\odot$ year$^{-1}$ at a distance of $57\pm6$ pc from the SMBH. For the first time, we calculate spatially resolved gas evacuation timescales for the cold molecular gas reservoirs ostensibly sourcing the outflows, and find that evacuating gas to $\sim$400 pc from the SMBH occurs on timescales of $10^{6.0} - 10^{7.6}$ years. These results indicate that the multi-phase AGN outflows are effective in clearing the inner few hundred parsecs of NGC 3227's gas content on timescales that may set the AGN duty cycle of $10^5 - 10^8$ years.
\end{abstract}



\section{Introduction}

At the center of nearly every galaxy lies a supermassive black hole (SMBH). In the nearby Universe, a small fraction (5--10\%) of SMBHs are active galactic nuclei (AGN), which are actively accreting surrounding gas and facilitating a complicated feedback process between the SMBH and its host galaxy. This dynamic occurs as enormous amounts of energy are released from the accretion process, resulting in outflows of gas that can evacuate reservoirs of potential star-forming gas from the galactic bulges \citep{ciotti01, heckman14, piotrowska21, booth09, angles17} and potentially suppress star formation within the galaxy \citep{fischer17, fischer18, revalski21, venturi21}. 

\begin{figure}[t]
\centering  
\includegraphics[width=0.9\linewidth]{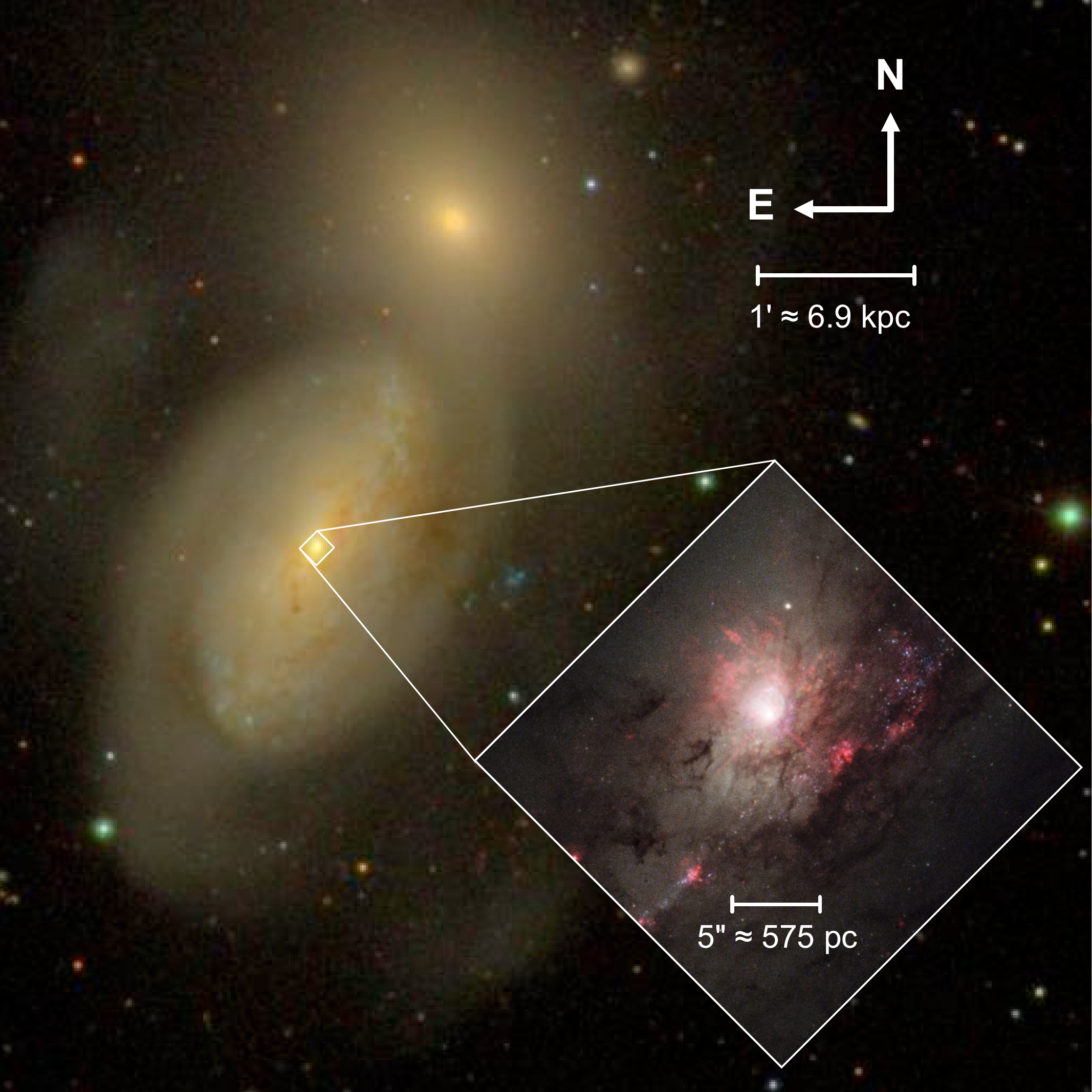}
\caption{Optical image of NGC~3227 (left of center) and NGC~3226 (above center) taken by the Sloan Digital Sky Survey with \textit{ugriz} filters. The inset shows a color-composite image of the inner region surrounding NGC 3227's AGN, where red colors are from the HST WFC3/UVIS F658N filter, green colors are from the HST WFC3/UVIS F547M filter, and blue colors show F550M and F330W filters from Hubble's Advanced Camera for Surveys High Resolution Channel. Image credit: NASA / ESA / Judy Schmidt.} 
\label{fig: 3227}
\end{figure}

A category of outflow known as AGN winds operate across a wide range of spatial (sub-parsec to kiloparsec) scales and different gas phases. The propagation of these winds has been extensively studied \citep{antonucci85, pedlar93, nelson00, pogge88, travisthesis}. Geometric and kinematic models show that on large scales, the ionized gas travels along a biconical geometry with the central vertex of the bicone coinciding with the AGN \citep{crenshaw00a, crenshaw00b}. 

To study the feedback mechanisms operating within galaxies, we perform studies on Seyfert galaxies \citep{seyfert43}, which are nearby ($z \leq 0.1$), and contain moderate-luminosity AGN ($L_{bol} \approx 10^{43} - 10^{45}$ erg s$^{-1}$).  Within Seyfert galaxies, we can study the impact of outflows and feedback by focusing on the kinematics of the narrow emission line region (NLR), which is composed of ionized gas at distances 1--1000 pc from the SMBH and has hydrogen densities ranging from $n_\mathrm{H} \approx 10^2 - 10^6$ cm$^{-3}$ \citep{revalski22}.  NLR outflows are a form of AGN winds that result from direct ionization and removal of reservoirs of cold molecular gas in the central regions of the galaxy, which would otherwise be available for star formation or fueling the SMBH \citep{storchi10, muller11, travisthesis, fischer17, king15, meena23}. 

Spatially resolved studies of NLR outflows are crucial because they connect the processes at the smallest scales (parsecs from the SMBH) to those on the scale of the galactic bulge (kiloparsecs) and beyond, and effectively lay the foundation for mechanisms that facilitate AGN feedback processes and affect galaxy evolution \citep{okamoto05, croton06, angles17}.
Specifically, the gas is likely pushed away from the nucleus via radiative driving \citep{proga00, das07, ramirez12, meena21}, in which the AGN-induced radiative acceleration and gravitational deceleration from the galaxy and SMBH control the velocity and physical extent of the NLR outflows.  

However, to more deeply understand the interplay between the outflows and the host galaxy, it is necessary to analyze not only the gas kinematics, but also the mass outflow rates that detail how the gas is evacuated by the outflows and subsequently distributed into the interstellar medium (ISM) \citep{dallagnol21, davies20, esposito24, revalski18a, revalski18b, revalski21, revalski22, trindade21}. The spatially resolved mass outflow rate directly addresses this question by quantifying the feedback as a function of distance from the SMBH, ultimately revealing crucial information such as the rates and timescales upon which this gas is evacuated from the NLR. With this information, we can characterize the effectiveness of the outflows in evacuating the cold gas reservoirs situated within the NLR, which informs us about the properties of the AGN phase lifetime (also known as the duty cycle).

Historically, studies of Seyfert galaxies have found a wide range of mass outflow rates and energetics over the spatial extent of the NLR \citep{barbosa09, storchi10, riffel09, muller11}. However, as a result of limited spatial resolution, these studies have predominantly produced global mass outflow rates represented by single (or occasionally a few) values. These studies are beneficial in that their results can be obtained relatively quickly; however, they rely on assumptions about the gas density and distribution that can significantly overestimate the outflow rates \citep{karouzos16, bischetti17}. To accurately account for the spatial distribution of the ionized gas, we utilize high-resolution ($<$0\farcs3) observations to calculate spatially resolved mass profiles and mass outflow rates. 



Recent studies of spatially resolved mass outflow rates for ionized and molecular gas have uncovered radial trends in the distributions of the mass, kinetic energy, and their outflow rates \citep{garcia14, crenshaw15, morgianti15, alonso19, bischetti19, zanchettin21, ramos22, revalski18b,revalski21, revalski22, revalski25, marconcini25}. However, most work on spatially resolved mass outflow rates has focused on a single gas phase (typically ionized or molecular), which provides an incomplete picture of the multiphase AGN winds \citep{cicone18}. With a multiphase, multiscale study, we can better understand the effectiveness of the feedback mechanisms based on how the various phases compare with one another throughout the NLR \citep{fluetsch19, shimizu19, garcia21, ramos22, zanchettin23, travascio24, speranza24, esposito24}.

This paper continues the work of \cite{falcone24}, hereafter referred to as Paper I. In Paper I, we performed a kinematic analysis on the ionized, neutral, and warm molecular gas in the Seyfert 1 galaxy NGC~3227 ($z$ = 0.003859), shown in Figure \ref{fig: 3227}. This nearby ($D = 23.7 \pm 2.6$ Mpc; \citealp{tonry01, blakeslee01}), SAB(s)a-type galaxy \citep{devaucouleurs91} has a weakly-barred structure with an interior spiral and strong evidence of interaction with its dwarf elliptical companion NGC~3226 \citep{rubin68, mundell95}. We used the ionized kinematics to determine the orientation of the outflowing bicone and conclude that radiative driving is the dominant acceleration mechanism for the NLR outflows in NGC~3227. 
This paper utilizes those findings to develop spatially-resolved mass and mass outflow rate profiles for three gas phases in NGC 3227, from which we will estimate the rate at which gas evacuation from the nuclear region occurs.





\section{Observations}
\label{sec: obs}
We use a combination of photometric and spectroscopic data with high ($\le$ 0\farcs 3) spatial resolution to map the ionized, warm molecular, and cold molecular gas. To characterize the ionized gas, we utilize the Hubble Space Telescope (HST) Wide Field Camera 3 (WFC3) to analyze [O~III]~$\lambda5007$ images that map the entirety of the NLR \citep{amanda24, marinelli24}. We use a F547M image, which has a spectral range of 5060–-5885 \AA, as the continuum to subtract from the line emission of the F502N image, which has a spectral range of 4969–-5044 \AA. Both images were obtained from HST program ID 16246 (PI: M. Revalski). We complement these photometric data with spectroscopy we obtained from the Kitt Peak Ohio State Multi-Object Spectrograph (KOSMOS) on the ARC 3.5-meter telescope at Apache Point Observatory,which has an observed 3$\sigma$ sensitivity of $\sim 3.48 \times 10^{-17}$ erg s$^{-1}$ cm$^{-2}$ \AA $^{-1}$. Further details on the WFC3 and KOSMOS data and reduction techniques, along with those from HST's Space Telescope Imaging Spectrograph (STIS) that we used for developing the kinematic models, are provided in Paper I. 

\begin{figure*}
\centering 
\subfigure{\includegraphics[width=0.42\linewidth]{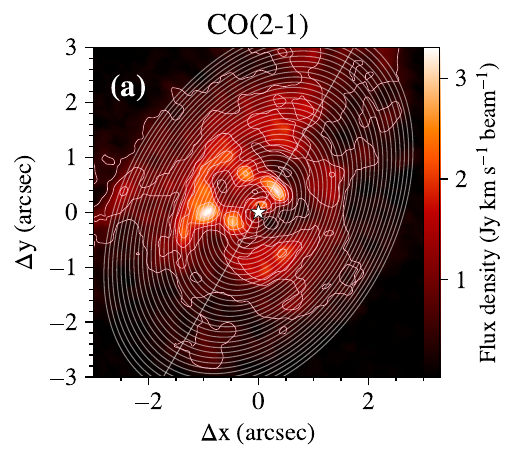}\label{fig: ALMA annuli}}
\subfigure{\includegraphics[width=0.39\linewidth]{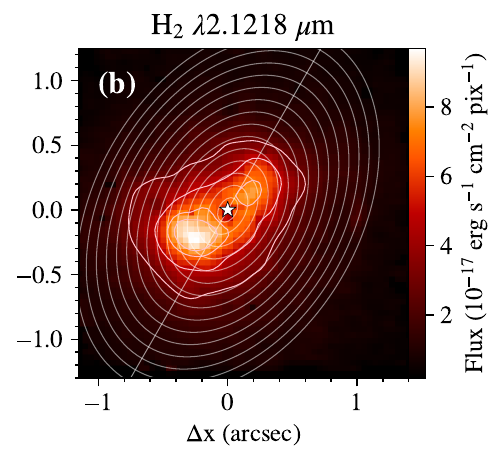}\label{fig: NIFS annuli}} 
\subfigure{\includegraphics[width=0.5\linewidth]{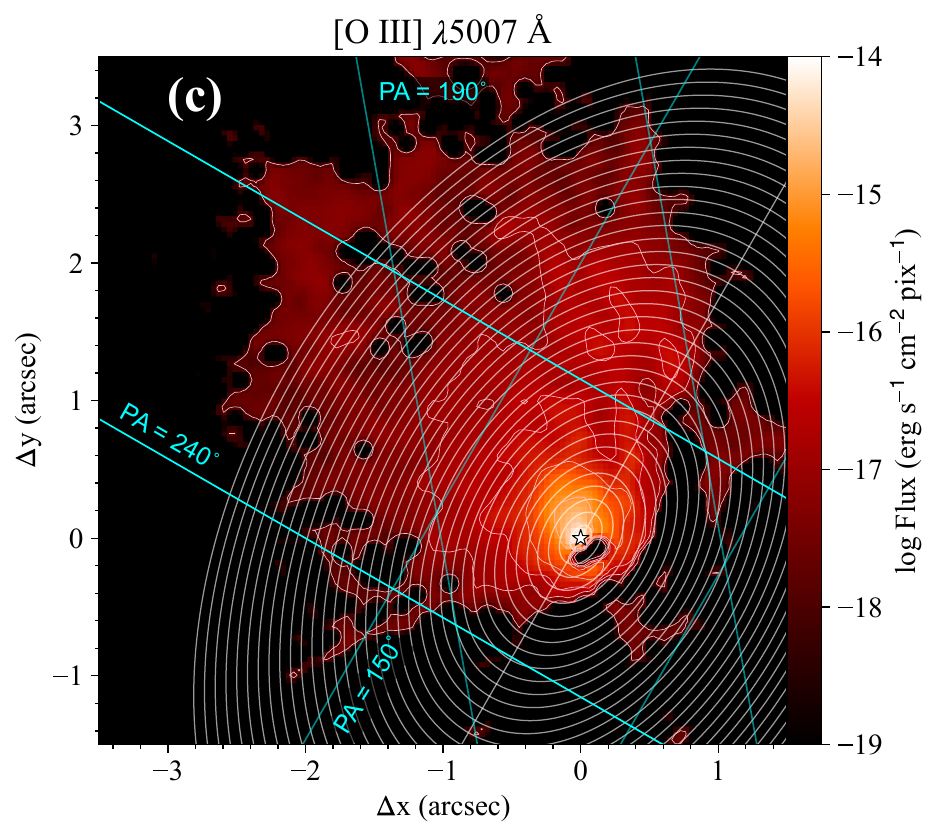}\label{fig: WFC3 annuli}}
 \caption{Multiphase flux maps of the central regions in NGC~3227. In each map, north is up and east is to the left, and the white star marks the position of the SMBH. (a) A map of the ALMA CO(2-1) flux density, originally shown in \cite{alonso19}. (b) A map of the NIFS H$_2$~$\lambda2.1218$ $\mu$m emission. (c) A map of a continuum-subtracted HST WFC3 F502N image of the nucleus, which shows the small-scale structure of the [O~III]~$\lambda5007$ emission. The heavy cyan lines represent the outline of the KOSMOS slit oriented along the galactic minor axis, which we use in our analysis. The light solid lines represent the KOSMOS slits oriented along the galactic major (PA = 150\arcdeg) and outflow (PA = 190\arcdeg) axes. In all three plots, the maps are overlaid with annuli that we describe in Section \ref{sec: annuli} and contours that better reveal the morphology of the flux distributions.}
 \label{fig: annuli}

 \end{figure*}

Figure \ref{fig: WFC3 annuli} shows the flux distribution for the ionized gas using the [O~III] emission from the WFC3 image, which has a measured 3$\sigma$ sensitivity of $ \sim2.52 \times 10^{-18}$ erg s$^{-1}$ pix$^{-1}$. In this study, we use the KOSMOS slit that runs along the minor axis  of the host galaxy (PA = 240\textdegree). We choose this orientation because of the thick dust lane in the galactic disk (see Figure \ref{fig: 3227}), which intersects the NLR outflows roughly along the major axis of the galaxy, and provides a clear divide between the obscured emission to the SW of the AGN and the mostly unobscured emission to the NE of the AGN.  

The disk of the galaxy is inclined to us at an angle of $i=$ 48\arcdeg \citep{devaucouleurs91, ho97again, xilouris02}. The position angle (PA) of the galactic disk is 150\arcdeg\ (Paper I), which we choose to be the major axes of our annuli. We use Equations 1 and 2 from \cite{revalski18b} to deproject the distance and velocity of the gas along these axes. With our adopted orientation ($i=48$\arcdeg, $\phi = 90$\arcdeg), our intrinsic distances are $\approx$~1.5 times larger than the observed distances along the minor axis. Thus, although we observe emission lines that meet our signal-to-noise (S/N) threshold of S/N $\ge$ 3 out to 2\farcs57 along the minor axis, which equates to 296 pc using a scale factor of 115 pc arcsec$^{-1}$ on the plane of the sky, the deprojected distance is 442 pc.


We consider the H$_2$ molecule to be representative of the molecular gas population due to its dominant abundance in the ISM. Although H$_2$ is not directly emissive at radio or sub-mm wavelengths, detectable transitions occur in near-IR wavelengths, including a strong emission-line at $\lambda2.1218$~$\mu$m. To characterize the warm molecular gas, we use archival observations of H$_2$~$\lambda2.1218$ $\mu$m emission from the \textit{K}-band of the Near-infrared Field Spectrograph (NIFS) at Gemini North (Program ID: GN-2016A-Q-6), which possesses a measured 3$\sigma$ sensitivity of $\sim3.78 \times 10^{-18}$ erg s$^{-1}$ cm$^{-2}$ \AA $^{-1}$. An in-depth description of these data is given in \cite{riffel17}, and the reduction processes of these data are described in Paper I. Each NIFS data cube covers $3\arcsec \times 3\arcsec$, which equates to an area of $\sim 345$~pc $\times $ 345 pc.

Because cold H$_2$ is not directly detectable in emission, we employ CO measurements as a tracer (see the review by \citealp{bolatto13}). To characterize the cold molecular gas, we use the ALMA CO(2-1) observations and associated calibrations described in detail in \cite{alonso19}. These observations, part of project 2016.1.00254S (PI: A. Alonso-Herrero), were obtained with band 6 at frequencies of the CO(2-1) transition (229.8 GHz) and the submillimeter continuum ($\sim$231 GHz, or 1.3 mm) at a bandwidth of 1.875 GHz \citep{alonso19} and with an observed 3$\sigma$ sensitivity of $\sim3.7 \times 10^{-4}$ Jy km s$^{-1}$ beam$^{-1}$. The resulting reduction yields a data cube with a beam size of 0\farcs161 $\times$ 0\farcs214, corresponding to a physical resolution of 15.5 $\times$ 24.6 pc. Although the field of view for these measurements are $\sim26\arcsec$, the relevant data are largely contained within a 7$\arcsec$ radius.

\section{Spectroscopic Analysis}
In Paper I, we measured the kinematics of the ionized gas in the NLR of NGC~3227 using multi-component Gaussian fits of the H$\alpha$~$\lambda$6563 $+$ [N~II]~$\lambda\lambda$6548, 6583 and H$\beta$~$\lambda$4861 $+$ [O~III]~$\lambda\lambda$4959, 5007 emission lines in both KOSMOS and STIS long-slit spectra. We used these measurements to disentangle the components of rotation and outflows as a function of distance from the SMBH. In this paper, we expand our spectroscopic analysis by using the observed line ratios to determine sources of ionization along the KOSMOS slits and develop an extinction correction.

\begin{figure*}[t]


\includegraphics[width=0.35\linewidth]{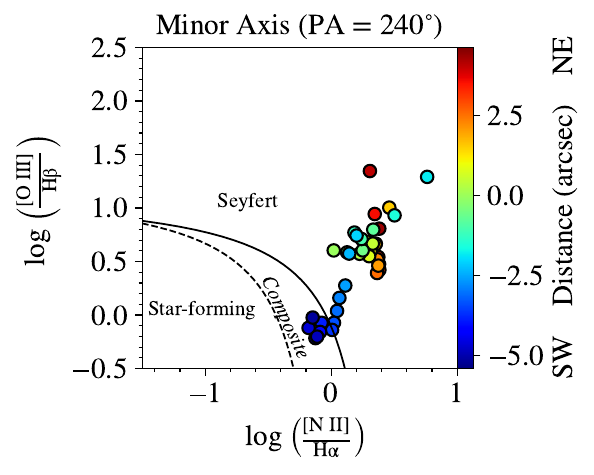}
\includegraphics[width=0.32\linewidth]{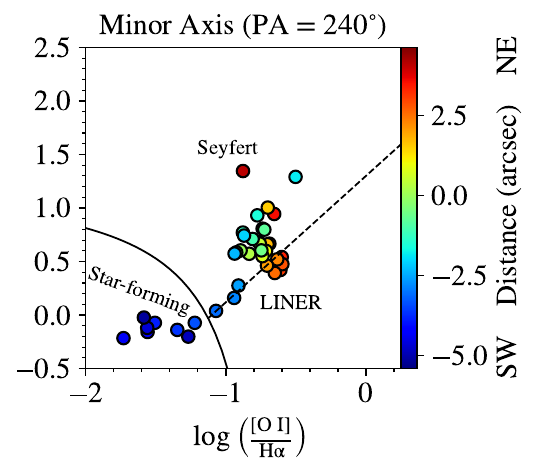}
\includegraphics[width=0.32\linewidth]{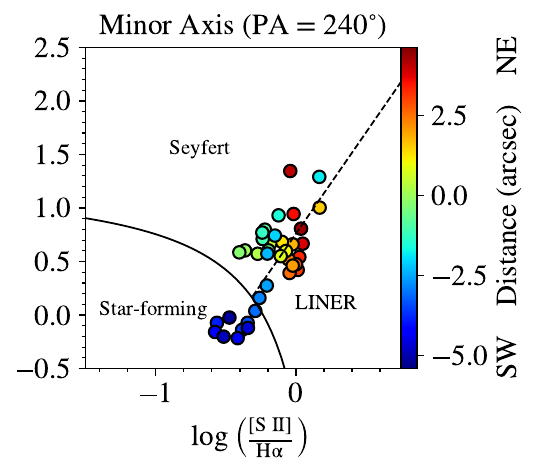}
 
\caption{BPT ionization diagrams for APO KOSMOS observations along the minor (PA = 240$^\circ$) axis. Positive distance refers to the north.}
\label{fig: BPT plots}
\end{figure*}



\subsection{Emission Line Fitting}
\label{sec: fitting}

We use two tools to perform the spectral fitting on our data. First, we fit multi-component Gaussian profiles using the Bayesian Evidence Analysis Tool (BEAT; \citealp{fischer17}) routine. BEAT is a novel tool in its use of Bayesian statistics to algorithmically determine the optimal number of kinematic components for a given spectrum. Because each spectrum contains the composite kinematic elements from the rotation of the galaxy and the outflowing motions of the winds, using BEAT to disentangle these components is critical in quantifying the kinematic contribution from the outflows. We used BEAT extensively in Paper I, and we use it again in this work to determine a radial velocity trend for the NIFS H$_2$~$\lambda2.1218$ $\mu$m emission. We do not use BEAT on the ALMA data because we were given access to the velocity map for these data (see \S \ref{sec: ALMA outflow rate}).



BEAT is optimized to fit only a few bright emission lines simultaneously, and those lines are typically close to each other in wavelength (such as H$\beta$ $\lambda$4861 and [O~III] $\lambda \lambda$4959, 5007, or H$\alpha$~$\lambda$6563 and [N~II]~$\lambda \lambda$6548, 6583). For fitting multiple emission lines simultaneously across a large wavelength range, we use an alternate procedure described in detail in \cite{revalski18a, revalski18b, revalski21, meena21} which aims to ensure that we fit the same kinematic components in each emission line for a given slit position.
This alternate routine uses BEAT fits of [O~III] $\lambda5007$ as a template to fit the narrow lines by adopting the velocities, positions, and widths of its components for the other lines while allowing the fluxes to vary to match the observed profiles. Nevertheless, they remain constrained by atomic line ratios such as those for the [N~II] $\lambda \lambda$6548, 6583 and [O~III] $\lambda \lambda$4959, 5007 doublets of 2.95 and 3.01, respectively \citep{osterbrock06}. We use this routine to fit the lines of H$\beta$~$\lambda$4861, H$\alpha$~$\lambda$6563, [N~II]~$\lambda \lambda$6548, 6583, [O~I]~$\lambda \lambda$6300, 6364, and [S~II]~$\lambda \lambda$6716, 6731.




\subsection{Ionization Source}
\label{sec: BPT}
One of the applications for myriad emission line measurements is to create Baldwin-Phillips-Terlevich (BPT) diagrams \citep{baldwin81, veilleux87} for NGC 3227. A BPT diagram is an important diagnostic tool that involves comparing line ratios of [O~III] $\lambda$5007/H$\beta$ to [N~II] $\lambda$6583/H$\alpha$, [O~I] $\lambda$6300/H$\alpha$, and [S~II] $\lambda$6730/H$\alpha$ to distinguish whether gas at a particular distance from the nucleus is ionized by the AGN, by star formation, or both.
We also use these emission-line fits to develop photoionization models for the AGN-ionized gas in the next section.

BPT diagrams from the KOSMOS spectra for the slit along the minor axis are shown in Figure \ref{fig: BPT plots}. The demarcations that separate the various classifications are described in \cite{kewley01, kewley06} and \cite{kauffmann03}.
The BPT diagrams for the other two KOSMOS slits are discussed in the Appendix.
Our BPT diagrams reveal a gradient in ionization as we move from the NE side of the slit to the SW. To the NE of the SMBH, the gas is primarily ionized by the AGN (``Seyfert''), but at $\approx3''$ the ionization levels weaken and the emission shows traits representative of a low-ionization nuclear emission region (LINER). This turn into the LINER designation is likely attributable to the weakening influence of the AGN at large distances, possibly due to radiation absorption by gas at nearer distances to the SMBH. On the southern side of the SMBH, the ionization level drops with distance from the SMBH. At --4$\arcsec$ from the SMBH, star formation starts to dominate over the AGN ionization. The area of prominent star formation coincides with a H~II region visible in the inset of Figure \ref{fig: 3227}. 
The thick dust lanes, which are more pronounced in the SE region of the nucleus, harbor star forming clumps visible in this figure, and also act to obscure AGN-ionized emission from the other side of the disk.



\subsection{Accounting for Extinction}
\label{sec: extinction}


The dust lanes at NGC 3227's center, which result in extensive reddening, have been an area of study for over four decades \citep{cohen83, winge95, crenshaw01, gondoin03, mehdipour21}. It is therefore important to  carefully consider how reddening effects impact the observed optical emission lines and their luminosities, as the luminosity measurements are critical to our calculations of the mass outflows. 



We can determine the reddening $E(B-V)$ according to the H$\alpha$/H$\beta$ ratio for a given spectrum. Specifically, we utilize Equation 3 of \cite{revalski18a}:
\begin{equation}\label{eq: E(B-V)}
E(B-V) \equiv  -\frac{2.5\text{ log}\left( \frac{F_o}{F_i} \right)}{R_\lambda} = \frac{2.5 \text{ log} \left(\frac{(H\alpha/H\beta)_i}{(H\alpha/H\beta)_o} \right)}{R_{H\alpha} -R_{H\beta} }
\end{equation}
where $F_o$ and $F_i$ are the observed and intrinsic fluxes, respectively, and $R_\lambda$ is the reddening value at a particular wavelength. We assume an intrinsic H$\alpha$/H$\beta$ ratio of 2.90, in accordance with recombination properties \citep{osterbrock06}.  From the extinction curve for NGC~3227 shown in \cite{crenshaw01}, we determine $R_{H\beta} \approx  $  3.67 and $R_{H\alpha} \approx$ 2.50. Thus, we have a direct relationship between the observed H$\alpha$/H$\beta$ ratio and the color excess. By employing Gaussian line fitting techniques (see \S \ref{sec: fitting}), we measure the H$\alpha$ and H$\beta$ fluxes to derive values for E(B-V).

Due to the strong presence of dust around the SMBH, the S/N of the H$\beta$ flux is drastically reduced to the extent that the reddening cannot be reliably determined at each location. \cite{schonell19} present a two-dimensional reddening map of NGC 3227 derived from Pa$\beta$/Br$\gamma$ line ratios obtained from NIFS data, but there are major spatial gaps in the data that cannot be accurately estimated through interpolation due to the clumpy nature of the data. Furthermore, low flux levels in the Pa $\beta$ and Br $\gamma$ emission to the NE and SW of the nucleus present uncertainty in the accuracy of their ratios.

We instead opt to estimate a value of $E(B-V)$ by summing all KOSMOS spectra along the minor axis within $\pm2''$ of the SMBH to create a composite spectrum, and perform a single fit of the H$\alpha$ and H$\beta$ fluxes to produce a global value for the reddening. This process yields a value of $E(B-V) = 1.07 \pm 0.12$. We apply this reddening value to all spectra, including those beyond $\pm2''$ because the intrinsic flux decreases significantly, resulting in regions of low S/N that are challenging to fit accurately and thus determine accurate reddening values in these regions. Our reddening value for the NLR of NGC~3227 is much larger than the value of $E(B-V) $= $ 0.18$ determined from the AGN continuum emission by \cite{crenshaw01}, which may indicate a smaller column of dust in the direct line of sight to the SMBH. Our reddening value of $E(B-V) = 1.07 \pm 0.12$  agrees well with the NLR values of \cite{schonell19} whose two-dimensional reddening map of NGC 3227 shows E(B-V) values near the central region in the range of 0.3 -- 2.2, and \cite{cohen83} who found E(B-V) = 0.94$~\pm~$0.23.

\section{Photoionization Models}

To calculate spatially resolved mass outflow rates using the techniques of \cite{revalski22}, we use the Cloudy code \citep{ferland13, chatzikos23}, which produces photoionization models for a given range of input parameters. This section describes our methodology to identify trends in these parameters as a function of distance, constrained by the emission-line fluxes, which allows us to generate models along the KOSMOS slit.

\subsection{Defining Extraction Annuli}
\label{sec: annuli}

Because we are calculating spatially resolved measurements of gas mass and mass outflow rate, we must extract fluxes from the images at equal distances along the inclined disks in Figure \ref{fig: annuli}. To do so, we utilize the Elliptical Panda routine within the SAOImage DS9 image software \citep{joye03}. We construct a series of concentric semi-ellipses centered on the nucleus \citep{revalski21}, with spacings equal to the spatial sampling of the HST spectra (2 pixels, or 0\farcs10156). The ellipticity of each annulus is calculated from our adopted inclination of 48\arcdeg\ \citep{xilouris02}. The ellipses are bisected along the major axis so that we collect data in bins across the minor axis (which is to say, from the SW to the NE).  We create these annuli for the flux maps for all three of our data sets, as shown in Figure \ref{fig: annuli}. Figures \ref{fig: ALMA annuli} and \ref{fig: WFC3 annuli} show that the annuli do not cover the full extent of either the CO(2-1) or [O~III] emission, and this is because the physical extent of our analysis is limited by our adopted gas velocity laws which do not extend past 400 pc (as described in Paper I). 


\subsection{Generating Cloudy Models}
\label{sec: cloudy parameters}

The ionization parameter $U$ is the dimensionless ratio of the number of ionizing photons to hydrogen atoms at the face of a gas cloud \citep{osterbrock06}.
$U$ and the hydrogen number density, $n_{\text{H}}$, are crucial parameters for the Cloudy models needed to determine the ionized gas mass of the cloud for a given AGN continuum spectral energy distribution (SED) and column density ($N_{\text{H}}$).

In our models, we choose an upper boundary of log($N_{\text{H}}$) = 24 cm$^{-2}$. The actual column of each model will be less than this value, and is established when the temperature drops to a value of 4000 K and the cloud becomes optically thick.

\cite{revalski22} show how the [O~III]/H$\beta$ emission line ratio varies primarily as a function of $U$ and secondarily with $n_{\text{H}}$. $U$ and $n_{\text{H}}$ are also related via the definition of the ionization parameter \citep{osterbrock06}:
\begin{equation}\label{eq: U_nH}
n_{\text{H}}=\left( \frac{Q(H)_{\text{ion}}}{4\pi r^2c \  U} \right)
\end{equation}
where $r$ is the radial distance of the gas from the SMBH, $c$ is the speed of light, and $Q(H)_{\text{ion}}$ is the number of ionizing photons s$^{-1}$ emitted by the AGN, given by $\int_{\nu_0}^{\infty}(L_\nu / h\nu)d\nu$, where $L_\nu$ is the luminosity of the AGN as a function of frequency (as denoted by the SED), $h$ is Planck's constant, and $\nu_0$ = 13.6 eV/$h$ is the ionization potential of hydrogen (\citealp{osterbrock06}, Section 14.3). We adopt a typical power-law SED that has been successfully utilized in previous studies \citep{kraemer00a, kraemer00b, kraemer09, revalski18a}. For $L_\nu \propto \nu^{\alpha}$, we choose slopes of $\alpha$ = --0.5 from 1 to 13.6 eV, $\alpha$ = --1.4 from 13.6 eV to 0.5 keV, $\alpha$ = ---1 from 0.5 to 10 keV, and $\alpha$ = --0.5 from 10 to 100 keV, with low- and high-energy cutoffs below 1 eV and above 100 keV, respectively.

We can therefore determine realistic values of $U$ and $n_{\text{H}}$ at each distance $r$ using the combined constraints of matching the observed [O~III]/H$\beta$ with photoionization models and satisfying Equation \ref{eq: U_nH}. 
In this study, we vary log($U$) from 0 to $-$4.2 in increments of 0.2, and for each value of log($U$) we also vary log($n_{\text{H}}$) from 1.5 to 4.7 cm$^{-3}$ in increments of 0.1.

\begin{figure*}[btp]
\centering  
\subfigure{\includegraphics[width=0.3\linewidth]{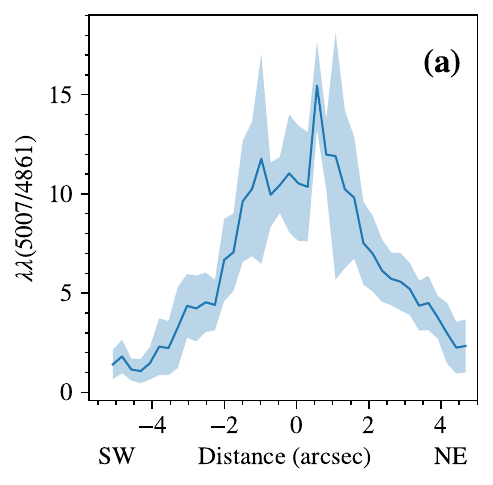}\label{fig: OIII-Hb kosmos}}
\hspace{2mm}
\subfigure{\includegraphics[width=0.305\linewidth]{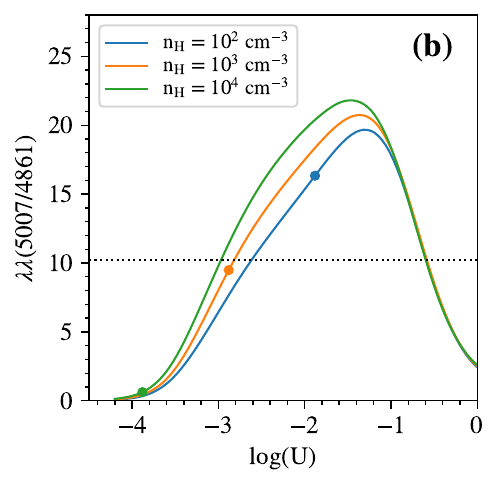}\label{fig: matching OIII-Hb}}
\hspace{2mm}
\subfigure{\includegraphics[width=0.325\linewidth]{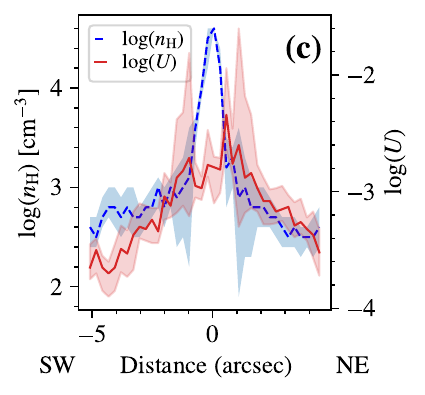}\label{fig: U and nH vs r}}
 \caption{(a) The observed [O~III]~$\lambda$5007/H$\beta$~$\lambda$4861 ratio as a function of distance along the kinematic minor axis, which has been corrected for projection effects. The shaded region represents the uncertainty in the observed line ratio.
 (b) The predicted [O~III]~$\lambda$5007/H$\beta$ ratio (curves) as a function of $U$ and $n_\mathrm{H}$ from a grid of Cloudy models. The point on each curve represents the $U$ and $n_\mathrm{H}$ that also satisfy Equation \ref{eq: U_nH} at a given distance (in this example, we chose 1\farcs34\ or 154 pc). The point that best matches the observed [O~III]/H$\beta$ ratio at this distance (dotted line) is selected as the correct $U$ and $n_{\text{H}}$ for this location. In practice, we use a finer grid of $U$ and $n_{\text{H}}$, only showing three curves here for illustrative purposes.
 (c) log($U$) (red, solid line) and log($n_\mathrm{H}$) (blue, dashed line) as functions of distance along the slit, where the shaded regions represent their respective uncertainties.}

\end{figure*}


We determine the [O~III]~$\lambda$5007 and H$\beta$~$\lambda$4861 emission line fluxes for each spectrum along our KOSMOS slit, which has a spatial scale of 0.257 arcsec pixel$^{-1}$, or 29.5 parsecs pixel$^{-1}$. To obtain these fluxes, we fit the spectra using the Gaussian line fitting techniques described in \S \ref{sec: fitting}. We show the [O~III]/H$\beta$ emission line ratio as a function of distance in our KOSMOS slit in Figure \ref{fig: OIII-Hb kosmos}. This ratio decreases with increasing distance from the SMBH, consistent with the trend of decreasing AGN ionization seen in the BPT diagrams (Figure \ref{fig: BPT plots}).

By utilizing the information from Figure \ref{fig: OIII-Hb kosmos}, we can determine the $U$-$n_{\text{H}}$ pair whose model's corresponding [O~III]/H$\beta$ ratio most closely matches that given by KOSMOS for a particular distance, as shown in Figure \ref{fig: matching OIII-Hb}. If we choose to look at a distance of +1\farcs34 as an example, we first use Figure  \ref{fig: OIII-Hb kosmos} to find the observed [O~III]/H$\beta$ ratio at that distance. That value is represented in Figure \ref{fig: matching OIII-Hb}  as the dashed line. 

To use Equation \ref{eq: U_nH} as a constraint we estimate $Q(H)_{\text{ion}}$ by taking the AGN continuum luminosity of Mrk~78, which is comparable to that of NGC~3227, and scaling it according to the relative ratio of bolometric luminosities between NGC~3227 and Mrk~78.  For Mrk~78 we use $L_{bol} = 7.9\times 10^{45}$  erg s$^{-1}$ and $Q(H)_{ion} = 3.8 \times 10^{54}$ photons s$^{-1}$ \citep{revalski21}, and for NGC 3227 we use $L_{bol} = 2.25\times 10^{44}$ erg s$^{-1}$ (Paper I). This results in $Q(H)_{ion} = 1.1 \times 10^{53}$ photons s$^{-1}$ for NGC~3227, which is applied to all distances from the nucleus.

In Figure \ref{fig: matching OIII-Hb}, we plot curves showing the Cloudy-predicted [O~III]/H$\beta$ ratios as a function of $U$ for three different densities. The points on these curves represent the $U$-$n_{\text{H}}$ pairs that satisfy Equation~\ref{eq: U_nH} at the chosen distance.
The $U$-$n_{\text{H}}$ pair that gives the lowest residual between the modeled and observed [O~III]/H$\beta$ ratio is chosen. 
We also note how Figure \ref{fig: matching OIII-Hb} shows that, regardless of the density, there is a degeneracy in the [O~III]/H$\beta$ ratios for log($U$) $\ge -1$. We therefore exclude any points at those $U$ values, consistent with our detection of strong low-ionization lines that would not be present at high values of $U$. In the example of Figure \ref{fig: matching OIII-Hb} where we are only choosing between three values of $n_{\text{H}}$, the orange curve which shows $n_{\text{H}}$ = 10$^3$ cm$^{-3}$ and log(U) = $-$2.88 would be considered the optimal model for this location. In practice, we use a finer grid as described above.

Figure \ref{fig: U and nH vs r} shows the results after completing this analysis for each spectrum along the KOSMOS slit. The plot shows how $n_H$ and $U$ vary with each other as a function of distance, with errors calculated according to uncertainties in the observed [O~III]/H$\beta$ ratios. The residuals  between the observed [O~III]/H$\beta$ ratios and the best models at each distance are low, usually only 0.1--0.4. The density is highest at the nucleus, reaching log($n_{\text{H}}$) $\approx 4.5$ cm$^{-3}$, and then drops off steeply on either side in a nearly symmetrical manner. A similar pattern is observed in the $U$ distribution as well, although its peak is at +0.5\arcsec NE, possibly due to the [O~III]/H$\beta$ ratio in Figure \ref{fig: OIII-Hb kosmos} peaking around that distance. We can compare these trends to those in \cite{revalski22}, whose study of six Seyfert galaxies shows trends in $U$ and $n_{\text{H}}$ that are also generally symmetric with densities decreasing with distance from the SMBH.
We choose to create Cloudy models, which assume a slab geometry, at evenly spaced distances separated by 0\farcs10156, as described in \S \ref{sec: annuli}. We create models in both the NE and SW directions, moving along the KOSMOS slit oriented on the minor axis as shown in Figure \ref{fig: WFC3 annuli}. We interpolate along the trends in [O III]/H$\beta$, \textit{U}, and $n_{\text{H}}$ as a function of distance as shown in Figures \ref{fig: OIII-Hb kosmos} and \ref{fig: U and nH vs r}, which allow us to generate unique parameters for each model. We create models extending to the edge of the bicone at $\sim$400 pc, where the velocity law goes to zero. We use the solar abundances as given in \cite{grevesse10}, which is a stored abundance set in Cloudy.





\section{Mass Outflow Rates and Evacuation Timescales: Techniques}

From the Cloudy models, we can determine mass outflow properties by extracting the model H$\beta$ model fluxes to use alongside our observed [O~III] fluxes. Although we divide our bins into hemispheric annuli as shown in Figure \ref{fig: annuli}, we are primarily concerned with radial trends under a general assumption of axial symmetry. Thus, in our subsequent calculations of mass and mass outflow trends, we perform the calculations for each semi-annulus and sum azimuthally.

The mass outflow rate ($\dot{M}_{\mathrm{out}}$) for each of our three phases can be calculated via
\begin{equation}\label{eq: Mdot}
\dot{M}_{\mathrm{out}} = \frac{Mv}{\delta r}
\end{equation}
where $M$ is the mass in each annulus, $v$ is the deprojected velocity which has been corrected for the effects of inclination and position angle, and $\delta r$ is the deprojected width of each annulus. The deprojection factor is the same for the warm and cold molecular gas as for the ionized gas because the available evidence indicates that the molecular disk lies in the plane of the galaxy \citep{alonso19}. Although the mass and velocity for each gas phase are dependent on several different factors, the deprojected width of each annulus is a constant value of $\delta r$ for all gas phases of 0\farcs102. The following subsections describe the methodology utilized to obtain the mass and velocity measurements for each of the three phases.

%

%
\subsection{Cold Molecular Gas }
\subsubsection{Mass Calculation}
To convert the CO(2-1) flux to H$_2$ mass, we follow \cite{alonso19}, who referred to Equation 2 of \cite{sakamoto99}:
\begin{equation} \label{eq: sakamoto}
\begin{split}
\left( \frac{M_{\mathrm{H}_2}}{M_\odot} \right) = 1.18 \times 10^4 \times \left( \frac{D}{\mathrm{Mpc}} \right)^2 \left( \frac{S_{\mathrm{CO(1-0)}}}{\text{Jy km s}^{-1}} \right)\\
 \times \left[ \frac{X}{3.0 \times 10^{20} \text{ cm}^2 \text{(K km s}^{-1})^{-1}} \right]
\end{split}
\end{equation}
where $D$ is the distance, $S_{\mathrm{CO}(1-0)}$ is the total CO(1-0) line flux, and $X$ is the CO-to-H$_2$ conversion factor. \cite{alonso19} estimate a brightness temperature ratio, $R_{21}$ = CO(1–0)/CO(2–1) = 1, based on an average value for spiral galaxies. For the CO-to-H$_2$ conversion factor, they use a value of $X = 2 \times 10^{20}$ cm$^{-2}$ (K km s$^{-1})^{-1}$ \citep{bolatto13}.  

We repeat this process, but we choose a conversion factor described in \cite{sandstrom13}, who perform a spatially-resolved study on the CO-to-H$_2$ conversion factor with local spiral galaxies and assume $R_{21}$ = 0.7. We choose this value over the one given in \cite{bolatto13} because the latter is ideal for galaxies like the Milky Way, whereas the former is a comprehensive study of $X_{\mathrm{CO}}$ in extragalactic sources. \cite{sandstrom13} solve for the conversion factor using the formula $X_{\mathrm{CO}}  = \alpha_{\mathrm{CO}} \times (4.6 \times 10^{19})$, where $ \alpha_{\mathrm{CO}}$ relates the CO flux to the H$_2$ surface brightness.
\cite{sandstrom13} find that the $\alpha_{\mathrm{CO}}$ profile is generally flat past $0.2r_{25}$ at a value of 3.1, where $r_{25}$ is the B-band isophotal radius at 25 mag arcsec$^{-2}$. At distances closer than $0.2r_{25}$, values for $ \alpha_{\mathrm{CO}}$ decrease significantly.

In NGC 3227, the value of r$_{25}$ is $91\farcs57 \pm 18\farcs72$ \citep{robinson21}. Our maximum distance of 2\farcs8 is only 2.7\% of $r_{25}$. \cite{sandstrom13} direct that for distances closer than 0.1$\times r_{25}$, the average $\alpha_{\mathrm{CO}}$ is a factor of two lower than the value of 3.1 associated with the remainder of the radius. Thus, when we factor that into our calculations, we find $X_{\mathrm{CO}} = 7.19 \times 10^{19}$ cm$^{-2}$ (K km s$^{-1}$)~$^{-1}$. 



\subsubsection{Mass Outflow Rate Calculation}
\label{sec: ALMA outflow rate}
To estimate the velocities associated with the CO(2-1) emission, we replicate the process of creating a rotation-subtracted CO(2-1) velocity field as described in detail in \cite{alonso19}. To isolate the outflows and noncircular kinematics in NGC 3227, \cite{alonso19} modeled the CO(2-1) data with $^{\mathrm{3D}}$BAROLO \citep{diteodoro}, which approximates the rotating galactic disk, and created a CO(2-1) residual mean velocity map by subtracting the modeled rotational kinematics from the CO(2-1) mean velocity field. For our study, we manually recreated the $^{\mathrm{3D}}$BAROLO model and subtracted the kinematics from the CO(2-1) mean velocity field to produce our own velocity residual plot, which is a nearly identical copy to the version shown in Figure 7 of \cite{alonso19}. 

Subtracting the $^{\mathrm{3D}}$BAROLO model yields a map of residual kinematics. These kinematics may represent outflows, or they may be associated with other local processes such as streaming motions or bar funneling. \cite{alonso19} has shown outflows to be dominant in the innermost 0\farcs2, and to be present alongside rotation out to 0\farcs5. Based on the observed kinematics and the assumption of a decelerating velocity law (as we have employed in our ionized data), we expect the outflows to extend in some form out to distances of $1''$.  Because of uncertainties in the origins of these motions, our analysis of the cold molecular gas represents \textit{maximum} possible values in the outflow properties. We hope to incorporate a spectral decomposition of these data, which would help disentangle the kinematic sources, into future analyses of NGC~3227. 

In our calculation of the maximum mass outflow rate, we interpret the residual kinematics as exclusively outflowing material  although later in our analysis, we will also assume the other extreme that none of the cold molecular gas near the nucleus is outflowing. In their analysis, \cite{alonso19} find contributions from structures such as a large-scale stellar bar to be insufficient to drive gas to the observed locations in position-velocity (p-v) diagrams. Rather, they find the kinematics can be most accurately replicated when with a model uses the $^{\mathrm{3D}}$BAROLO model as a base and adds a noncircular radial velocity component (i.e., outflows) and a nuclear warp in the galactic disk. However, their model without the warp produces similar results in its ability to reach specific locations in the p-v diagrams. Therefore, although \cite{alonso19} do not find outflowing motions to exclusively define the kinematics, our assumption is reasonable for the determining the maximum outflow rate.

\begin{figure}[t]
\centering  
\subfigure{\includegraphics[width=\linewidth]{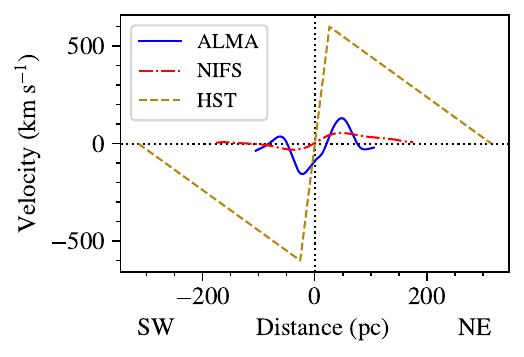}}

\caption{The velocity laws for the ALMA CO(2-1) (blue solid line), NIFS H$_2$ $\lambda2.1218$ $\mu$m (red dot-dashed line), and HST [O~III] $\lambda 5007$ (gold dashed line) data that are used to calculate the maximum mass outflow rates for the cold molecular and warm molecular gas, and the approximate mass outflow rate for the ionized gas.}
\label{fig: ALMA and NIFS velocity laws}
\end{figure}

However, there is a systemic velocity of $\sim$ 30 km s$^{-1}$ that remains in both our CO(2-1) velocity residual and the version in \cite{alonso19}. This systemic velocity causes an offset in the velocity field which creates a substantial asymmetry in the residual velocities. The cause of this systemic velocity may be attributable to inflowing streaming motions resulting from a large-scale stellar bar \citep{davies14}, which correlate with expected kinematics for H$_2$ and CO(2-1) emission \citep{alonso19}. Thus, in order to further isolate the residual kinematics from other sources of motion near the nucleus, we subtract an additional 30 km s$^{-1}$ from the residual velocity map to produce the version we use in our analysis. The resulting velocity law for the CO(2-1) data is shown in Figure~\ref{fig: ALMA and NIFS velocity laws}.

\subsection{Warm Molecular Gas}
\subsubsection{Mass Calculation}


To calculate the gas mass from the H$_2$ $\lambda2.1218$ $\mu$m emission in the NIFS data cube, we employ Equation~6 from \cite{storchi09}:
\begin{equation} \label{eq: storchi H2 mass}
\begin{split}
M_{\mathrm{H_{2}}}  &= \frac{2m_\mathrm{p} \ F_{\mathrm{H}{_\lambda}2.1218} \ 4 \pi D^2}{f_{\nu=1, J=3}A_{S(1)} h\nu}\\
 & = 5.0776 \times 10^{13} \left( \frac{F_{\mathrm{H}{_\lambda}2.1218}}{\mathrm{erg \ s}^{-1} \mathrm{ \ cm}^{-2}} \right) \left( \frac{D}{\mathrm{Mpc}} \right)^2 
\end{split}
\end{equation}
where $m_\mathrm{p}$ is the proton mass, F$_{\mathrm{H}{_\lambda}2.1218}$ is the line flux, $D$ is the distance to the galaxy, $f_{\nu=1, J=3}$ is the population fraction, $A_{S(1)}$ is the transition probability, and the resulting $M_{\mathrm{H}_2}$ is in units of solar masses. \cite{storchi09} assume a vibrational temperature of $T_{\mathrm{vib}}$ = 2000 K, yielding $f_{\nu=1, J=3} = 1.022 \times 10^{-2} $ and $A_{S(1)} = 3.47 \times 10^{-7}$ s$^{-1}$. The orientation of our semi-annuli relative to the flux density of the H$_2$ $\lambda2.1218$ $\mu$m emission is shown in Figure \ref{fig: NIFS annuli}.

\subsubsection{Mass Outflow Rate Calculation}
We operate under the same assumption described in \S \ref{sec: ALMA outflow rate} that we are computing the \textit{maximum} warm molecular mass outflow rate. 

The rotation-subtracted velocity map of these H$_2$ data are shown in Figure 10 of Paper I. We recreate the process of overlaying the semi-annuli on this map in the same manner shown in Figure \ref{fig: NIFS annuli}, and taking the average velocity in each of those wedges. The resulting velocity law is shown in Figure \ref{fig: ALMA and NIFS velocity laws}.
Interestingly, both cold and warm molecular gas phases show velocity turnovers at similar radii (40 -- 60 pc) as the ionized gas phase, although with much smaller velocity amplitudes and much quicker returns to zero velocity at $\sim$100 pc.
We employ Equation~\ref{eq: Mdot} to solve for the outflow rate as a function of distance.

\subsection{Ionized Gas Mass}
\label{sec: ionized mass}
\subsubsection{Mass Calculation}
We calculate the luminosities of our [O~III] $\lambda 5007$ semi-annuli through the formula
\begin{equation}\label{eq: luminosity}
L(\lambda 5007) = 4 \pi D^2 F_{\lambda 5007} \times 10^{0.4 \times E(B-V) \times R_{5007}}
\end{equation}
where \textit{D} is the distance to the galaxy, $ F_{\lambda 5007}$ is the intrinsic [O~III] $\lambda5007$ flux given by the HST WFC3 data, and $R_{5007}$ is the reddening value for $\lambda$5007. \S \ref{sec: extinction} describes our methodology to determine values for $E(B-V)$ and $R_{5007}$. We use constant values of $R_{5007} = 3.65$ and $E(B-V) =1.07 \pm 0.12$. 

To convert $L_{5007}$ to a mass, we utilize the equation \citep{peterson97, crenshaw15}:

\begin{equation}\label{eq: Mslit}
M = N_{\mathrm{H}} \mu m_p \left( \frac{L(H\beta)}{F(H\beta)_m}\right)
\end{equation}
where $N_\mathrm{H}$ is the model hydrogen column density (which we obtain from the Cloudy results), $\mu$ is the mean mass per particle (which is 1.40 for solar abundances), $m_p$ is the mass of a proton, $L(H\beta)$ is the luminosity of H$\beta$ that we obtain from our conversion of the [O~III] luminosity, and $F(H\beta)_m$ is the H$\beta$ model flux that we obtain from the Cloudy models. We convert $L(\lambda 5007)$ to $L(H\beta)$ at each annulus by utilizing our interpolation of the [O~III]/H$\beta$ flux ratios shown in Figure \ref{fig: OIII-Hb kosmos}.  

\subsubsection{Mass Outflow Rate Calculation}
As shown in Paper I (see Figure 13 in that paper), the ionized gas in NGC~3227 is completely dominated by outflow to a distance of at least 400 pc from the SMBH.
To calculate the ionized mass outflow rate, we employ Equation \ref{eq: Mdot} where the mass at each distance is calculated in the previous subsection. \cite{travisthesis} show that Seyfert NLRs often follow an empirical velocity trend, wherein the velocity profile starts at $\sim$0 km s$^{-1}$ in the nucleus and increases roughly linearly until it reaches a turnover radius, at which point the velocity declines until it approaches systemic at the full extent of the NLR. In Paper I, we found that the NLR of NGC~3227 follows this trend and we used extensive modeling to determine that the turnover radius for NGC 3227 is 26$^{+6}_{-6}$ pc, at which location the velocity reaches a maximum of 600 km s$^{-1}$.  After reaching this peak, the velocity linearly declines until it reaches the full extent of the bicone at $\sim$400 pc, as shown in Figure \ref{fig: ALMA and NIFS velocity laws}. To calculate the ionized mass outflow rate, we interpolate at each distance from our velocity law.

\subsection{Timescale Calculation}
\subsubsection{Depletion Timescale}
\label{sec: depletion}
Assuming that the cold molecular gas reservoir revealed in the CO(2-1) data shown in Figure \ref{fig: ALMA annuli} is the source of this galaxy's AGN outflows, we can use the mass outflow rates to calculate the time it will take to evacuate the reservoir. 
In determining an evacuation timescale ($t_e$) for moving the gas to a fixed overall distance, we must consider both the depletion timescale ($t_d$) for the cold molecular gas to be removed and accelerated within its original annulus, and the crossing timescale ($t_c$) for gas in each phase to move across the remaining annuli, such that $t_e = t_d +t_c$.

For the purposes of this work, we are assuming a static environment in which gas is not driven inwards. This inflow rate has not been quantified for NGC 3227, but simulations have shown nuclear inflow rates to be highly variable and difficult to accurately quantify on our scales \citep{bournaud11, gabor13}. Thus, we opt to exclude the impact of inflows from our study, but future analyses of mass outflows may wish to consider this aspect.

When considering how gas is propagated outwards, we have shown that most of the outflowing gas is created in situ, likely from radiation pressure on the local dusty molecular gas \citep[Paper I]{das07, fischer17, meena21, meena23}. Thus, we consider two extremes when making our calculations: 1) all of the outflowing gas continues to be pushed outwards from one annulus to the next, or 2) none of the outflowing gas in an annulus is found in subsequent outer annuli. Physically, the latter situation could occur if the gas is either ionized to a higher phase in the annulus, and therefore undetectable in the visible, if it is driven to lower outflow velocities at larger radii and is not decelerating, or if it decelerates and does not reach the next annulus. The second and third cases are likely happening to some extent because the mass outflow rate declines significantly at distances past the peak in this and other AGN \citep{revalski21}. Some portion of the first case is likely occurring as well for the ionized gas, because our radiative driving analysis in Paper I shows that high-velocity ($>$ 200 km~s$^{-1}$) ionized clouds at locations $>$ 100 pc in NGC~3227 were launched from distances $<$ 10 pc from the SMBH, indicating movement across many annuli. This scenario is likely true for the cold and warm molecular gas outflows as well. Thus, the real situation is likely between these two extremes, which we can use to calculate limits on the timescales.

\begin{deluxetable*}{lccccc}
\vspace{-0.5em}
\setlength{\tabcolsep}{0.18in}
\def\arraystretch{0.95}
\tablecaption{Properties of the Mass Outflow Rates}
\tablehead{
\colhead{Gas Phase} & \colhead{$\dot{M}_{\mathrm{peak}}$} & \colhead{$\dot{M}_{\mathrm{peak}}$ Distance} & \colhead{Extent of Outflows} &\colhead{Integrated Gas Mass \vspace{-.6em}} \\
\colhead{} & \colhead{(M$_\odot$ yr$^{-1}$)} & \colhead{(pc)} & \colhead{(pc)} &\colhead{(M$_\odot$) \vspace{-1.5em}} \\
}
\startdata
Cold Molecular & $18.9 \pm 4.2$&  57 $\pm$ 6&  $92 \pm 6$& $(2.213 \pm 0.001) \times 10^8$\\ 
Warm Molecular& $(6 \pm 3)\times10^{-4}$ &  36 $\pm$ 6&  $164 \pm 6$& $(1.63 \pm 0.22) \times 10^3$\\ 
Ionized & $19.9 \pm 9.2$  &  47 $\pm$ 6&  $423 \pm 6$& $(7.7 \pm 0.6) \times 10^6$\\
\enddata
\tablecomments{A summary of the findings for each of the three phases studied in this work. The columns list (1) gas phase, (2) peak mass outflow rate for the azimuthally summed profiles (3) deprojected distance from the SMBH at which the peak mass outflow rate occurs, (4) total distance over which outflows are observed, and (5) the integrated gas mass up to the extents in column 4. \vspace{-1.7em}} 
\label{table: results}
\end{deluxetable*}

The depletion timescale $t_d$ for an environment where none of the outflowing gas moves to subsequent annuli, and therefore all created locally, can be simply calculated from $t_d = M_{\mathrm{H_2}} / \dot{M}_{\mathrm{out, \ total}}$, where $M_{\mathrm{H_2}}$ is the molecular mass for a given bin assuming that the ionization time scale in the inner regions of the galaxy is comparatively small and can be neglected \citep[see, e.g.][]{peterson13}, and $\dot{M}_{\mathrm{out, \ total}}$ is the sum of the individual mass outflow rates for the cold molecular, warm molecular, and ionized phases within a given bin. In this calculation, we further divide $\dot{M}_{\mathrm{out, \ total}}$ into two options: a maximum value, which incorporates the maximum mass outflow rates for the cold and warm molecular gas, and a minimum value, wherein the cold and warm molecular gas outflow rates are set to zero and the only contribution comes from the ionized gas outflows.

In situations where all of the gas is pushed outward to the next annulus, it is necessary to define the net ionized mass outflow rate for each phase, $\dot{M}_{\mathrm{net}}$,  as $\Delta \dot{M}_{\mathrm{out}}$ from one annulus to the next moving outward. This is because the outflowing gas mass being pushed forward from a previous annulus must still be carried forward in subsequent annuli, and so it must be considered when discussing the capacity to evacuate the gas.
The depletion timescale in this case is given by $t_d = M_{\mathrm{H_2}} / \dot{M}_{\mathrm{net, \ total}}$, where 
\begin{align*}
\dot{M}_{\mathrm{net, \ total}} = \dot{M}_{\mathrm{net,\ ion.}} + \dot{M}_{\mathrm{net, \ warm \ mol.}} + \dot{M}_{\mathrm{net, \ cold \ mol.}}. 
\end{align*}
We also calculate $\dot{M}_{\mathrm{net, \ total}}$ as a maximum and minimum value in the same way as described above for $\dot{M}_{\mathrm{out, \ total}}$. $\dot{M}_{\mathrm{net, \ total}}$ is not relevant and therefore not calculated when $\dot{M}_{\mathrm{out}}$ declines from one annulus to the next, because no molecular gas is being removed from that annulus.




\subsubsection{Crossing \& Evacuation Timescales}
\label{sec: evacuation}
The time that is required to remove the gas to some outward boundary is known as the crossing timescale. Specifically, we define the crossing timescale $t_c$ as the time for the gas of a particular phase at a given distance to be pushed from that distance to the expected extent of the outflows, which we limit according to the extent of our velocity laws as shown in Figure \ref{fig: ALMA and NIFS velocity laws}. The crossing timescale for every individual annulus of each phase is calculated from $\Delta t_c = M / \dot{M}_{\mathrm{out}}$ ($ = \Delta r / v$), where $M$ is the gas mass of the given phase within that annulus and $ \dot{M}_{\mathrm{out}}$ is the outflow rate of the given phase within that annulus. The values are then cumulatively added starting from the edge of the outflows and moving inwards to that annulus. Each gas phase has its own crossing timescale according to its specific velocity law.

Finally, the crossing timescale is added to the depletion timescale to calculate the evacuation timescale $t_e$ for each phase at every annulus. The evacuation timescale describes the total amount of time for gas to be removed from the reservoir and expelled from within the inner regions surrounding the SMBH, extending $100 - 400$ pc depending on the phase.

It should be noted that although the topic of gas evacuation timescales has been the subject of much study \citep{cicone14, fiore17}, our work takes a slightly different approach because of our novel spatially resolved methodology. As a result, other works such as \cite{cicone14} use the term ``depletion timescale'' to refer to what we call the evacuation timescale because they rely on a single global value to describe the overall time required to evacuate the entire nuclear region.



\subsection{Sources of Uncertainty}

\subsubsection{Cold Molecular Gas}
\label{sec: molecular errors}
The primary source of random error that we consider in the cold molecular gas mass estimate is the uncertainty in the CO(2-1) flux measurements. We estimate the uncertainty by first designating a continuum region where there is minimal emission on the flux map. The uncertainty is calculated with the equation

\begin{equation}
\sigma_{\text{line}} = \sqrt{\frac{f_{\text{line}}}{f_{\text{cont.}}}} \times \sigma_{\text{cont.}}
\label{eq: unc}
\end{equation}
where $f_{\text{line}}$ is the average line flux per pixel, $f_{\text{cont.}}$ is the average continuum flux per pixel, and $\sigma_{\text{cont.}}$ is the standard deviation in the continuum region region.
We find that $\sigma_{\text{line}} \approx 0.00035$ Jy km s$^{-1}$ beam$^{-1}$, which yields a fractional uncertainty of $< 1$\% after propagation. We repeat this process on the velocity field to obtain an uncertainty in the mass outflow rates, and obtain a fractional uncertainty of $\sim1.5$\%. 

Additionally, there are potential systematic errors that depend on choices of conversion factors and can be used to scale our results accordingly.
The primary source of systematic uncertainty is the CO-to-H$_2$ conversion factor, $\alpha_{\mathrm{CO}}$, which can vary by as widely as 0.3~dex in the centers of galaxies \citep{sandstrom13}.  A few of the main factors contributing to this uncertainty pertains to weaknesses in the correlation between $\alpha_{\mathrm{CO}}$ and metallicity; uncertainty in $R_{21}$, the CO(2-1)/CO(1-0) brightness temperature ratio; and biases related to quantifying the dust-to-gas ratio. 
\cite{sandstrom13} estimate 0.2~dex for the error of $R_{21}$. This estimation is in line with that presented by \cite{braine92}, who find $R_{21}$ = 0.78 for NGC 3227 as part of a survey carried out using the IRAM 30m telescope. 
Another potential systematic error is that of the distance to NGC 3227, which we take to be $D = 23.7 \pm 2.6$ Mpc \citep{tonry01, blakeslee01}.
We exclude systematic uncertainties from our plots with the understanding that future improvements could be used to scale our results accordingly.

\subsubsection{Warm Molecular Gas}
The errors in the warm H$_2$ mass result primarily from uncertainties in the fluxes measured in the NIFS data. As in the previous subsection, we use Equation~\ref{eq: unc} to estimate the uncertainty using the line fluxes and continuum region emission of the H$_2$ data, resulting in errors that range from 50--70\%.

In calculating the uncertainty in the mass outflow rate, we divide our H$_2$ velocity map into the same semi-annuli shown in Figure \ref{fig: NIFS annuli}, and find the velocity error for each annulus by taking the standard deviation of the velocity measurements within it.

\subsubsection{Ionized Gas}
The dominant source of error in the ionized gas mass and outflow rate calculations is the uncertainty in our estimate of the reddening, $E(B-V)$, which is due in most part to uncertainties associated with fitting the observed emission line fluxes for H$\alpha$ and H$\beta$ to determine a value for $E(B-V)$ as described in Section \ref{sec: extinction}. The errors in the fits are quite small, especially because the fitted spectrum covering $\pm 2''$ represents the sum of multiple spectra, which decreases the noise significantly. However, these values are compounded into large uncertainties for $L(\lambda5007)$ (see Equation \ref{eq: luminosity}).

The uncertainty in the extinction also contributes to the uncertainty in the conversion from [O III] $\lambda$5007 to H$\beta$ $\lambda$4861, which is needed to determine Cloudy input parameters (see Section \ref{sec: cloudy parameters}). In order to perform this conversion, we utilize the empirical $\lambda \lambda$5007/4861 relationship as a function of distance from the nucleus, which we obtain by comparing the emission line ratios of [O III] $\lambda$5007 and H$\beta$ $\lambda$4861 in the KOSMOS data (see Figure \ref{fig: OIII-Hb kosmos}). 

\section{Mass Outflow Rates and Evacuation Timescales: Results}
\label{sec: results}

\subsection{Mass Profiles and Outflow Rates}
\begin{figure*}[t]
\centering  

\subfigure{\includegraphics[width=0.47\linewidth]{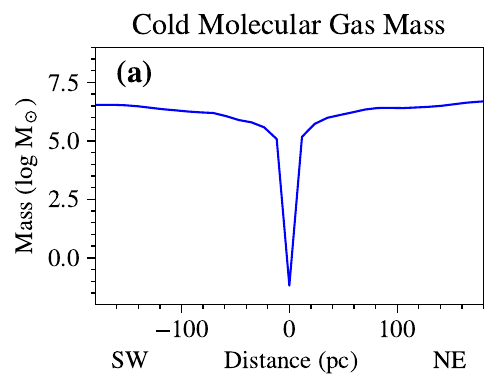}\label{fig: ALMA mass}}
\subfigure{\includegraphics[width=0.48\linewidth]{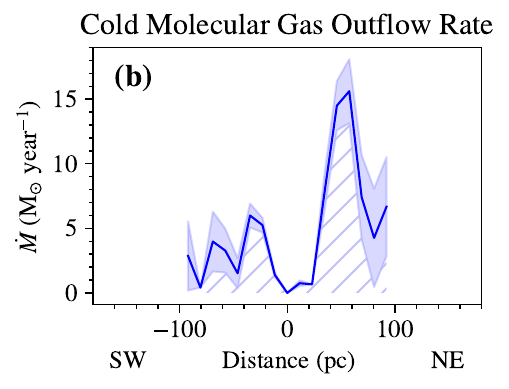}\label{fig: ALMA outflow rate}}
\vspace{-.2cm}

\subfigure{\includegraphics[width=0.45\linewidth]{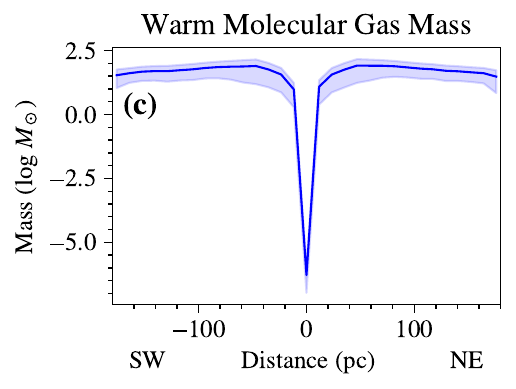}\label{fig: NIFS H2 mass}}
\subfigure{\includegraphics[width=0.51\linewidth]{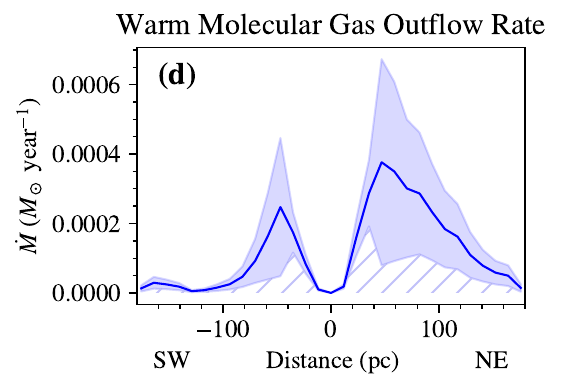}\label{fig: NIFS H2 outflow rate}}
\vspace{-.2cm}

\subfigure{\includegraphics[width=0.47\linewidth]{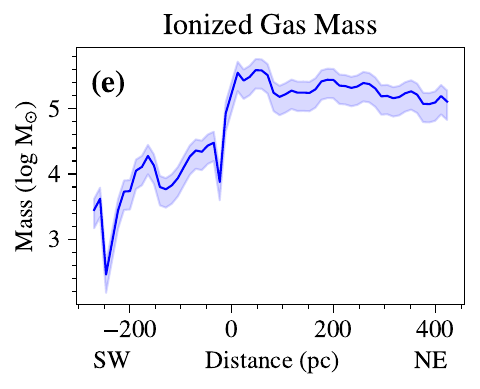}\label{fig: OIII mass}}
\subfigure{\includegraphics[width=0.48\linewidth]{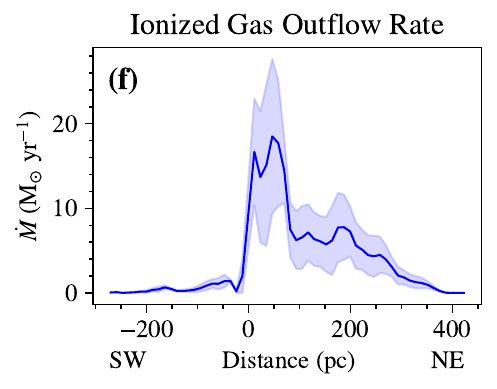}\label{fig: OIII outflow rate}}
\vspace{-.2cm}

\caption{The gas mass (left) and mass outflow rate profiles (right) as a function of distance for each of the three gas phases: cold molecular (top), warm molecular (middle), and ionized (bottom).  The warm and cold molecular gas outflow rates represent upper limits and the hatch markings represent the scope of our uncertainty, whereas the ionized gas rate represents actual values. Trends in the mass and outflow rate profiles are generally symmetric around the nucleus for the warm and cold molecular gas, but show strong asymmetries for the ionized gas due to significant extinction. 
}
\label{fig: mass and outflow profiles}
\end{figure*}

Figure \ref{fig: mass and outflow profiles} presents the spatially resolved mass and mass outflow rate profiles in the cold molecular, warm molecular, and ionized gas phases as a function of distance from the SMBH in NGC~3227 in bins of 11.5 pc, continuing under the assumption that the cold and warm molecular gas outflow rates represent upper limits. We include hatch markings for the outflow rates of the cold and warm molecular gas to indicate our uncertainty in the amount of gas present in the outflows.
The extents of these profiles are limited to those of the velocity profiles in Figure \ref{fig: ALMA and NIFS velocity laws}, which approach zero velocity in the rest frame of the galaxy at distances 100 -- 400 pc from the SMBH. The primary findings for the three phases are listed in Table \ref{table: results}.

\subsubsection{Cold and Warm Molecular Gas}
\label{sec: molecular results}

Figure \ref{fig: mass and outflow profiles} shows that the cold and warm molecular gas phases share similarities in the shapes and extents of their mass and maximum outflow rate profiles. The mass profiles rise sharply from the center before leveling out around $\sim$50 pc, reflecting the deficit of molecular emission seen near the nucleus in Figure \ref{fig: annuli}.
The dynamics between cold and warm molecular gas populations in the central regions of galaxies are still being understood, with some studies finding the warm H$_2$ acts as a thin shell or skin that surrounds the cold H$_2$ gas reservoir, heated by the AGN radiation \citep{storchi10, riffel21, bianchin22}, and other studies finding that the warm H$_2$ occupies cold molecular gas cavities \citep{rosario19, feruglio20}.  Our work under the assumption of observed outflows aligns with the former interpretation, and as such, we expect the mass of the warm H$_2$ emission to be magnitudes lower than that observed for the cold H$_2$ emission, as has been seen in other AGN \citep{dale05, muller06, bianchin22}. 
We find this to be true: Figure \ref{fig: mass and outflow profiles} and Table \ref{table: results} show that the warm molecular gas mass is $\sim$5 orders of magnitude lower than the cold gas mass. 

\begin{figure*}[t]
\centering  

\subfigure{\includegraphics[width=0.47\linewidth]{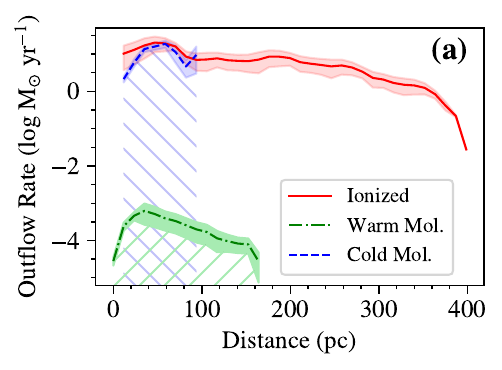}\label{fig: mass outflow rates plotted together}}
\hspace{3mm}
\subfigure{\includegraphics[width=0.47\linewidth]{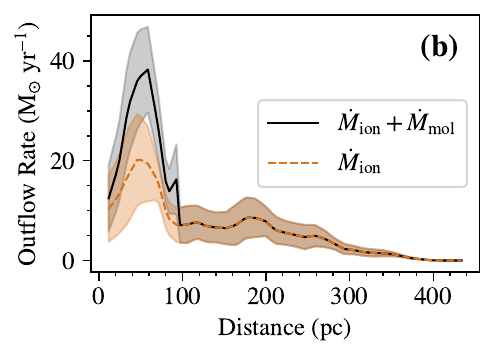}\label{fig: mass outflow rates phases combined}}

\caption{(a) The mass outflow rates from the three phases, as shown in Figure \ref{fig: mass and outflow profiles}, are plotted. We have plotted them as a function of radial distance from the center by summing the mass outflow rates at equal distances in either direction from the SMBH. The hatch markings represent the systemic uncertainty in estimating the contributions of the cold and warm molecular gas towards their outflows. (b) The composite mass outflow rates, comprising the sum of the ionized and molecular outflow rates from the left panel (black, solid curve) and only the ionized mass outflow rate (orange, dashed curve).}
\label{fig: composite mass outflow rate}
\end{figure*}


Using observations of CO and HCN with the IRAM Plateau de Bure Interferometer at 0\farcs6 resolution, \cite{schinnerer00} found that NGC~3227 hosts a molecular ring of gas centered around the SMBH with a diameter of 3$''$ (345 pc) and with stronger emission on its eastern side.
Subsequent studies \citep{hicks08, davies14, schonell19} have sought to elucidate the role of the ring on the nuclear kinematics of NGC~3227, generally by characterizing the noncircular motions. \cite{alonso19} are the first to claim that these noncircular motions imply gas outflows, based on their high-resolution ALMA observations of CO, and this idea is supported by our spatial analysis of the cold and warm molecular kinematics in Paper I. 

Based on this finding of outflows in the molecular ring, if we assume that the entirety of the ring may be outflowing, Figure \ref{fig: ALMA outflow rate} shows that we obtain maximum cold molecular gas outflows that peak at $\le 18.1$ M$_\odot$ yr$^{-1}$ at 57 $\pm$ 6~pc NE and $\le 6.9$ M$_\odot$ yr$^{-1}$ at 34 $\pm$ 6 pc SW of the SMBH. A discussion of uncertainties for these values is found in \S \ref{sec: molecular errors}. These values are substantially larger than the values of 5 M$_\odot$ yr$^{-1}$ and 0.6 M$_\odot$ yr$^{-1}$ for the NE and SW ends, respectively, calculated by \cite{alonso19}.  However, it should be noted that in calculating the outflow rate using Equation \ref{eq: Mdot},  \cite{alonso19} confined the outflows to a 0\farcs2 square aperture whereas  Figure \ref{fig: ALMA outflow rate}  shows outflows that peak outside of this aperture and extend to almost 1\arcsec. 

Figure \ref{fig: NIFS H2 outflow rate} shows that for the warm molecular gas, we see maximum outflow rates with one peak of $\dot{M}\mathrm{_{out}}\le 3.0 \times10^{-4}$ M$_\odot$ yr$^{-1}$ at a distance of 47 $\pm$ 6 pc SW of the nucleus, and another peak at $\dot{M}\mathrm{_{out}}\le 4.4 \times 10^{-4}$ M$_\odot$ yr$^{-1}$ at a distance of 47 $\pm$ 6 pc NE of the nucleus, which creates a combined maximum annular mass outflow rate of $\dot{M}\mathrm{_{out}}\le 7.4\times10^{-4}$ M$_\odot$ yr$^{-1}$ for this gas. The warm molecular gas mass and mass outflow rate are five orders of magnitude smaller than those of the cold molecular gas, which agrees with observations that the cold gas mass constitutes the majority of the gas mass in galaxies \citep{dale05, muller06}.  Warm molecular outflows are also detected in NGC~3227 by \cite{bianchin22}, whose study of the same NIFS H$_2$ $\lambda2.1218$ $\mu$m data reveals outflow rates of $\dot{M}\mathrm{_{out}}= 6.4 \times 10^{-4}$ M$_\odot$ yr$^{-1}$, which closely agrees with our values.

\subsubsection{Ionized Gas}
\label{sec: ionized results}
Figures \ref{fig: OIII mass} and \ref{fig: OIII outflow rate} show very significant ionized gas masses and mass outflow rates, but stark differences when comparing the NE region to the SW. This is due to the strong obscuration effects of dust, which are especially prominent in the SW (see the inset in Figure \ref{fig: 3227}). As such, the observed gas mass is likely undercounted, leading to an underestimate of the outflow rate in that region. Based on our model of the biconical outflows described in Paper I, we expect the outflow trends in the NE and SW to be fairly symmetric with one another. 

When the semi-ellipse annuli are azimuthally summed, our radial ionized mass outflow rate shown in Figure \ref{fig: mass outflow rates plotted together} reveals a trend that increases sharply to reach a peak of 19.9 $\pm$ 9.2 M$_\odot$ yr$^{-1}$ at 47 $\pm$ 6 pc, maintains an average value of 6  M$_\odot$ yr$^{-1}$ for the distance spanning $\sim$100--200 pc, and then steadily declines until it drops to 0 at $\sim 400$ pc. 

\begin{figure*}[t]
\centering  
\subfigure{\includegraphics[width=0.48\linewidth]{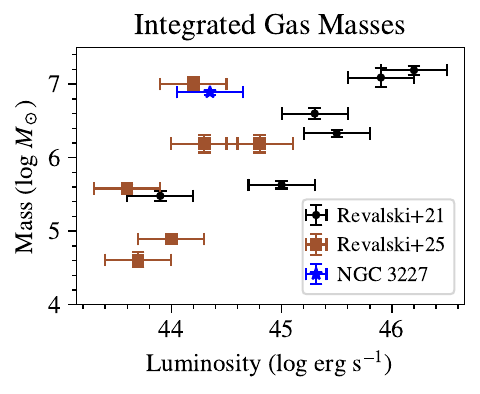}\label{fig: Mitch mass comparison}}
\hspace{3mm}
\subfigure{\includegraphics[width=0.49\linewidth]{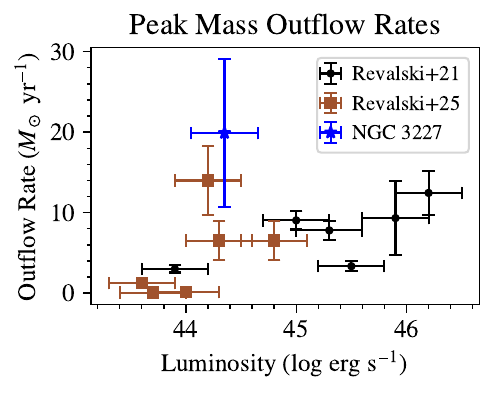}\label{fig: Mitch outflow comparison}}

\caption{The integrated ionized gas masses (left) and peak ionized outflow rates (right) of the AGN studied in \cite{revalski21} and \cite{revalski25}, alongside those for NGC~3227 presented in this work.}
\label{fig: Mitch comparisons}
\end{figure*}

In Figure \ref{fig: Mitch comparisons}, we compare our integrated ionized mass and mass outflow peak for NGC~3227 to those in Figure 5 of \cite{revalski25}, which also analyzes spatially-resolved emission to graph mass outflow rates for six galaxies alongside six galaxies previously studied in \cite{revalski21}. Our values for the peak ionized mass outflow rate and total integrated ionized mass are $\dot{M}$ = 19.9 $\pm$ 9.2 M$_\odot$ yr$^{-1}$ and log($M$) = $6.9 \pm 0.03$ M$_\odot$. We see that NGC~3227 generally follows the established trend in mass with bolometric luminosity, but it stands out as having the highest outflow rate of the sample despite a relatively modest bolometric luminosity (log$(L_{bol}) \sim  44.35$ erg s$^{-1}$; see Paper I). Nevertheless, its peak rate is close to that of NGC 788 at a similar luminosity. Similar positive trends connecting mass outflow rates to the bolometric luminosity have been previously studied \citep{cicone14, fluetsch19}, but the addition of spatially-resolved elements allows us to study these trends at a new level of detail.

\subsubsection{Composite Mass Outflows}
\label{sec: composite outflows}
Figure \ref{fig: mass outflow rates plotted together} plots the maximum mass outflow rates for the cold molecular, warm molecular, and ionized phases for complete annuli, where we use hatch markings to signify the systemic uncertainty related to the contributions of the cold and warm molecular gas towards the outflows. Our results show how the outflow rates of the cold molecular and ionized gas in this case are similar to each other, while the warm molecular gas outflow rate is more than four magnitudes lower.  

Studies have found that the cold molecular outflows tend to be larger than the ionized outflows by a factor of $10-10^3$ for $L_{AGN} \lesssim 10^{46}$ erg s$^{-1}$\citep{carniari15, fiore17, fluetsch21}, although others have found ionized outflow rates in galaxies that are 1--8 times larger than the molecular outflows \citep{venturi21}.  \cite{fluetsch19} find that for AGN with luminosities $L_{bol} \approx 10^{44}$ erg s$^{-1}$ like NGC~3227, molecular outflows are about an order of magnitude more powerful than the ionized outflows. Thus, NGC~3227 can be considered an outlier in this regard.


Figure \ref{fig: mass outflow rates phases combined} shows the sum of the mass outflow phases, added for each azimuthally-summed radial bin, plotted alongside the ionized mass outflow rate. In doing so, we can estimate the impact of the outflows at their highest (with the presence of molecular outflows) and their lowest (without molecular outflows) values. The combination of the ionized and molecular outflows leads to a peak outflow rate of $\dot{M}\mathrm{_{out}}= 37.9 \pm 8.4$ M$_\odot$ yr$^{-1}$ at $54 \pm 6$ pc, which afterwards falls off steeply past 100 pc, where outflows of the cold molecular gas are no longer detected. Due to the lack of published material regarding multiphase mass outflow rates, it is challenging to contextualize our composite outflow values. Nevertheless, \cite{fluetsch19} record several Seyfert galaxies for which global ionized and molecular mass outflow rates combine to reach 10$^2$ - 10$^3$ M$_\odot$ yr$^{-1}$, although the contributions from the molecular outflows dominate over those from the ionized outflows. 

\begin{figure*}[t]
\centering  
\subfigure{\includegraphics[width=0.465\linewidth]{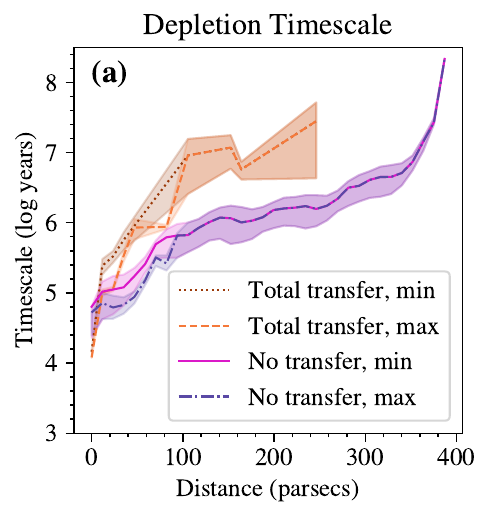}\label{fig: depletion timescale}}
\subfigure{\includegraphics[width=0.44\linewidth]{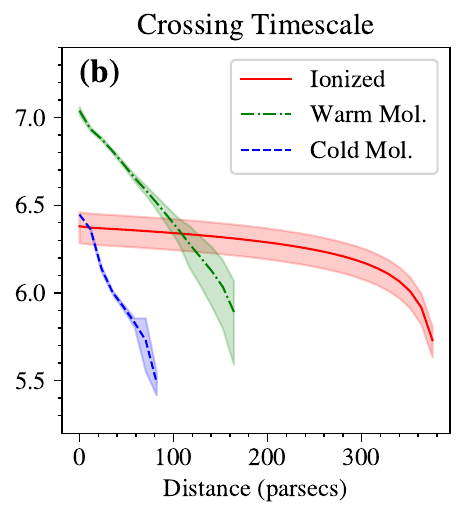}\label{fig: crossing time}}
\vspace{-.2cm}

\caption{(a) The depletion timescale for the molecular gas reservoir to be emptied by outflows in the three phases studied and (b) the crossing timescale for each of the three phases. The labels ``total transfer" and ``no transfer" refer to the extreme situations in which all or none of the gas mass is transferred from one bin to the next, respectively. The labels ``max'' and ``min'' refer to the presence and absence of warm and cold molecular outflows, respectively. The ``total transfer'' case has fewer points and a shorter maximum distance than the ``no transfer" because it can only be quantified for positive $\dot{M}_{\mathrm{net, \ total}}$ values at each distance.}
\label{fig: dep and cross timescales}
\end{figure*}

\begin{figure*}[t]
\centering  

\subfigure{\includegraphics[width=0.335\linewidth]{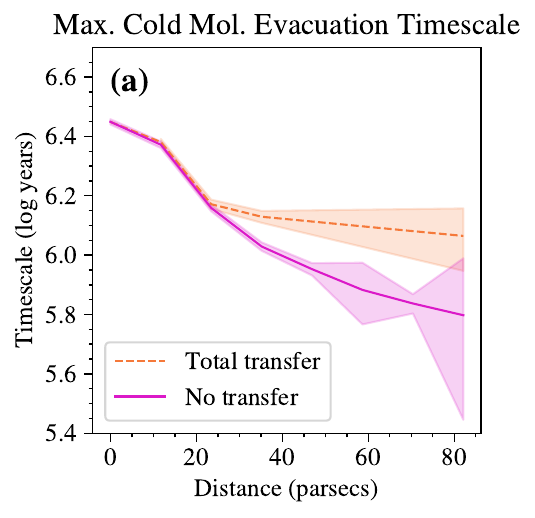}\label{fig: evacuation cold molecular}}
\subfigure{\includegraphics[width=0.32\linewidth]{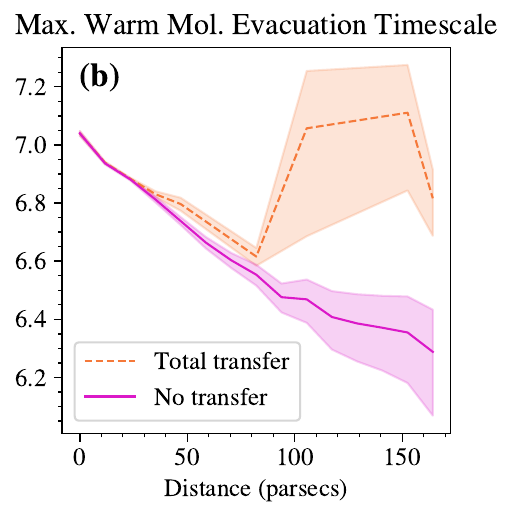}\label{fig: evacuation warm molecular}}
\subfigure{\includegraphics[width=0.31\linewidth]{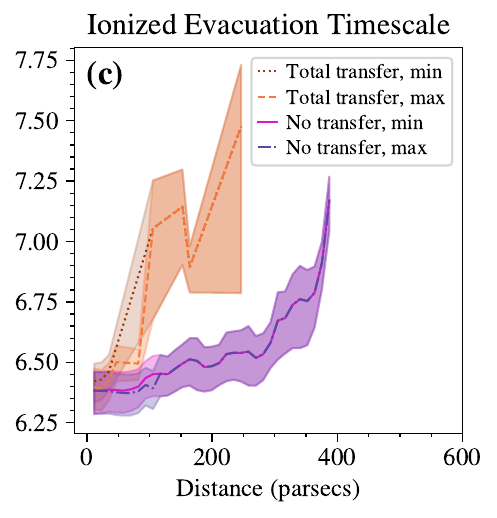}\label{fig: evacuation ionized}}

\caption{The evacuation timescales for (a) maximum cold molecular, (b) maximum warm molecular, and (c) ionized gas. The meanings of the ``total transfer," ``no transfer," ``min,'', and ``max'' labels are the same as in Figure \ref{fig: dep and cross timescales}.}
\label{fig: evac timescales}
\end{figure*}

\subsection{Resulting Timescales}

The depletion timescale as a function of distance from the SMBH can be calculated for all three phases combined from the cold molecular gas profile in Figure \ref{fig: ALMA mass} and the mass outflow rates in Figure \ref{fig: mass outflow rates phases combined}. This calculation is made for the two extremes described in Section \ref{sec: depletion} in which either all gas or no gas is propagated radially outwards from one annulus into the next. In Figure \ref{fig: depletion timescale}, the ``total transfer'' scenario yields higher timescales than the ``no transfer'' one because the net mass outflow rates $\dot{M}_{\mathrm{net}}$ are lower in the former. Within these two scenarios, we also note how the ``minimum'' cases, which exclude the molecular mass outflow rates from the calculations, result in higher timescales than the ``maximum'' cases, which include the molecular and ionized outflows. This is simply because without the molecular outflows contributing to the excavation of the reservoirs, it will take longer to deplete them. The depletion of the cold gas reservoir in each annulus occurs on timescales from $10^{4.5} - 10^{7.5}$ years, generally increasing with distance, with variations up to $\sim$1 dex depending on how much mass is transferred from one location to the next. It is also significant to note from Figure \ref{fig: depletion timescale} that the ionized outflows are the dominant drivers of outflowing gas in this AGN system, which sets it apart from the systems that are dominated by molecular outflows as discussed in the previous subsection.

Figure \ref{fig: crossing time} shows the crossing timescales, which depend on gas phase and span from $10^{5.3}$  -- $10^{7.0}$ years over the inner few hundred parsecs. As mentioned in \S \ref{sec: evacuation}, because the crossing timescale spans the distance from the current location to the edge of the outflows, $t_c$ will always increase as distance to the SMBH decreases. This figure shows that for the cold and warm molecular gas, $t_c$ increases significantly with decreasing distance from the SMBH. The crossing timescale of the ionized gas is fairly constant over most of the distance, ranging from $(2-5) \times 10^4$ years in each radial annulus until 350 pc, at which point the mass outflow rate declines significantly. 

Figure \ref{fig: evac timescales} shows the separate evacuation timescales for the three phases of gas. The cold and warm molecular evacuation timescales plotted represent the ``maximum'' cases, whereas the ``minimum'' cases of $\dot{M}_\text{mol}=0$ are not plotted. For the cold and warm molecular gas, In general, we observe timescales on the order of $10^{6.0} - 10^{7.6}$ years for the gas evacuation in the inner 400 pc to occur. The evacuation timescales for the warm and cold molecular gas are heavily influenced by longer crossing timescales at distances $<$ 50 pc, and by increasing depletion time scales at distances $>$ 50 pc. Because the crossing time for the ionized gas decreases slowly with distance, its evacuation timescales closely reflect the trends of its corresponding depletion timescales. The values at each annulus vary as widely as $\sim$1 dex between the ``total transfer'' and ``no transfer''  trends. Interestingly, the evacuation time scales span similar ranges (10$^6$ -- 10$^7$ years) in all three phases in the inner $\sim$100 pc, with the upper limit increasing to 10$^{7.6}$ years at larger distances for the ionized gas.

To contextualize these results, we can compare these timescales to the length of the AGN duty cycle. It should be noted that the literature is inconsistent when referring to the ``duty cycle'': in some contexts it refers to the total amount of time that an SMBH is active over its lifetime, while in other contexts it refers to the length of time in which an SMBH is continuously active, otherwise known as the ``AGN lifetime.'' In this work we adhere to the latter definition. 

Comparing our results to predictions of duty cycles is complicated because of large uncertainties. There are findings that the phase can last $10^7 - 10^9$ years \citep{martini01, marconi04}, but there are other suggestions that the longer phase is actually split into many shorter phases lasting $\sim 10^5$ years \citep{schawinski15, clavijo24}.  Furthermore, in certain situations some AGN-driven molecular outflows may continue to propagate $\sim10^8$ years after the AGN has turned off \citep{king11}. 

Nevertheless, our results show that removal of cold gas reservoirs due to mass outflows in the inner $\sim$400 pc of a moderate-luminosity Seyfert galaxy can be accomplished in  $10^{6.0} - 10^{7.6}$ years. This process can therefore be a mechanism for establishing the AGN duty cycle over these time scales, assuming no replenishment of cold gas during this time.
Other studies have identified AGN that may evacuate the molecular gas content within this time span \citep{sturm11, cicone14, fiore17, fluetsch19}.

\section{Discussion}

\subsection{NGC~3227's Cold Molecular Outflow Rates}

We can compare our current study to that done by \cite{esposito24}, who calculate spatially-resolved mass outflow rates for cold molecular and ionized gas in NGC 5506. This galaxy is a comparable analog to NGC~3227 because both galaxies are classified as Seyfert~1, are at relatively similar distance (21 Mpc and 26 Mpc for NGC~3227 and NGC 5506, respectively), and have similar bolometric luminosities ($\sim$2.25 $\times$ 10$^{44}$ erg~s$^{-1}$ and $\sim$1.3 $\times$ 10$^{44}$ erg s$^{-1}$ for NGC~3227 and NGC~5506, respectively). Using CO(3-2) measurements, they record a trend in the molecular gas mass outflow rate that peaks at $\dot{M}{_{\mathrm{out, max}} ^{\mathrm{mol}}}$ = 28 M$_\odot$ yr$^{-1}$ at a distance of 85 pc (the site of the galaxy's molecular ring) and decreases to single-digit values past 2$''$. Although our own CO(2-1) data does not extend that far, the similarities in scale and magnitude of our observed maximum molecular outflow trends are notable.

Our results also concur with those of \cite{ramos22}, who compile molecular mass outflow rates across a variety of AGN encompassing QSO2s, Seyferts, and ultra-luminous infrared galaxies (ULIRGs). Their Figure 18 shows a trend relating the outflow mass rate to the bolometric luminosity. Their outflow mass value for NGC~3227, which comes from \cite{alonso19}, falls far below the trend. However, if we compare the trend with the results presented in this work, where the molecular mass outflow rate is predominantly above 2 M$_\odot$ yr$^{-1}$ and peaks at 15.6 M$_\odot$ yr$^{-1}$, we see that NGC~3227 is much more agreeable with the other galaxies. This comparison further underscores the benefit of spatially resolved mass outflow analyses.

\subsection{NGC~3227's Elevated Ionized Outflow Rates}

NGC~3227 is unusual in its elevated ionized outflow rates, and in \S \ref{sec: composite outflows} we describe how the ionized outflow rates are similar to the maximum cold molecular outflow rates between 40 -- 80 pc, despite evidence that molecular outflow rates are typically magnitudes higher than ionized outflow rates. 
We wish to better understand why our cold molecular mass outflow rates agree with the sample presented by \cite{esposito24}, whereas our ionized outflow rates differ. Following our discussion in the previous subsection, \cite{esposito24} record an integrated ionized mass outflow rate of $0.076 \pm 0.017$ M$_\odot$ yr$^{-1}$ for NGC 5506, which is two orders of magnitude lower than our spatially-resolved rates for NGC~3227. The lower outflow rate likely occurs because NGC 5506 possesses an ionized mass of ${M} = 9.8 \times 10^{4}$ M$_\odot$, which is far lower than the ionized gas mass we find for NGC~3227 of ${M} = (7.7 \pm 0.6) \times 10^6$ M$_\odot$. However, the integrated cold molecular mass estimates are very similar: in this work we record a molecular mass of $2.2 \times 10^8$ M$_\odot$ for NGC~3227, and \cite{esposito24} record a molecular mass of $1.75 \times 10^8$ M$_\odot$ for NGC 5506. Thus, the ionized and molecular mass estimates may explain the corresponding mass outflow rates.

We also compare our results to those in \cite{bianchin22}, who find ionized outflows for the bicone model on the scale of 0.008 - 0.18 M$_\odot$ yr$^{-1}$. This is attributable to their calculation of the integrated ionized gas mass for NGC~3227, which they find to be $(0.19 - 3.72) \times 10^4$ M$_\odot$. This is over two orders of magnitude lower than our ionized gas mass of $(7.7 \pm 0.6) \times 10^6 $ M$_\odot$. Our mass estimate is likely higher because we factor the effects of reddening into the calculation, which increases the mass estimate tenfold. According to \cite{schonell19} and as shown in Figure \ref{fig: Mitch mass comparison}, ionized gas masses for Seyferts tend to be higher than $10^4$ M$_\odot$, and it is likely that NGC~3227 would follow this trend as well.

\subsection{Influence of Stellar Feedback?}
It is well documented that feedback processes resulting in ionization and gas driving can result from AGN feedback, which is the focus of this work, and star formation feedback \citep{silk98, melioli15, drummond17}. If we utilize a measure of the star formation rate (SFR) within NGC~3227 to better understand the extent to which gas may be expelled due to star formation-driven outflows, we can further constrain the impact of the observed AGN-driven outflows.

\cite{schonell19} has calculated the SFR for the inner $3'' \times 3''$ of NGC~3227 to be $(0.9 \pm 0.2) \times 10^{-3}$  M$_\odot$ yr$^{-1}$, arriving at this value through calculating the mass accretion rate and the SFR surface density using NIFS data of NGC~3227 \citep{riffel17}. However, as we have shown with our BPT diagrams in Figure \ref{fig: BPT plots}, the AGN is the predominant source of ionization in this area. This finding is further confirmed by the fact that analysis of spectra outside the NLR show minimal signs of stellar absorption, implying that even beyond the immediate vicinity of the nucleus, the AGN is the preeminent source of ionizing radiation. Thus the SFR surface density, which is calculated from NIFS measurements of the Pa$\beta$ emission line, is contaminated by the AGN source and is not a reliable proxy for the SFR. A more accurate galactic SFR could be estimated if measurements were taken far beyond the bounds of NGC~3227's NLR bicone, where AGN contamination would be minimized. Alternatively, other studies have used measurements of warm dust with Herschel \citep{sturm11} to determine an estimate of SFR, but no such observations for NGC~3227 exist. Additionally, spectral energy distribution (SED) modeling in the IR would also reveal the heating sources of the dust \citep{garcia22}. Future studies on the star formation properties in NGC~3227's nucleus might wish to pursue one of these methodologies to achieve an estimation that effectively filters out the interference from the AGN.

In the absence of available concrete measurements of the SFR on either galactic or nuclear scales, we can instead use other observations to obtain general estimates. Namely, \cite{riffel17} uses the same NIFS data as \cite{schonell19} to connect velocity dispersion maps to the presence of young stars. The fact that NGC~3227 displays a centrally-peaked velocity dispersion rather than patches of low velocity dispersion reveals a lack of young stars in NGC~3227's inner $3'' \times 3''$. 

All these factors considered, we can confidently say that the SFR of NGC~3227 is very small compared to its mass outflow rates. Indeed, for a spiral galaxy with a stellar mass of $1.1 \times 10^9$ M$_\odot$ \citep{mundell95}, the SFR we expect to observe is approximately two magnitudes lower than the peak outflow rate of 20 -- 40 M$_\odot$ yr$^{-1}$ \citep{renzini15}. 

\section{Conclusions}

We present spatially resolved gas mass and mass outflow rate profiles for NGC~3227 in the cold molecular, warm molecular, and ionized gas phases, with the warm and cold molecular rates representing upper limits in this system. 
This study joins a small but growing number of studies focusing on spatially resolved mass outflow rates \citep{revalski18b, revalski21, revalski22}, even fewer of which are multiphase \citep{esposito24}. 
Our conclusions are as follows:
\begin{enumerate}
    \item The maximum cold molecular, maximum warm molecular, and ionized gas show peak mass outflow rates of $23.1$ M$_\odot$ yr$^{-1}$, $9\times10^{-4}$ M$_\odot$ yr$^{-1}$, and $19.9 \pm 9.2$ M$_\odot$ yr$^{-1}$, respectively. In all three phases, the peaks occur 35 -- 60 pc from the nucleus. When summed, the maximum peak outflow rate is $37.9 \pm 8.4$ M$_\odot$ yr$^{-1}$ and occurs around $54 \pm 6$ pc, where bright knots of CO emission are located at the ends of a $\sim$1\arcsec\ molecular bridge that crosses the location of the SMBH.
    \item Enormous variations in the outflow rates as a function of distance (both radially and azimuthally) underscore the necessity of spatially-resolved trends rather than relying on single (i.e. global) values, particularly for understanding the locations and origins of the outflows and their role in depleting the gas reservoirs.
    \item The peak ionized outflow rate agrees well with the trends observed by \cite{revalski21, revalski25}, which establish positive correlations between bolometric luminosity and both ionized gas mass and peak ionized mass outflow rate.
    \item We find the ionized mass outflow rates and maximum cold molecular in NGC~3227 to be nearly equal with one another, which is in contrast with studies of most other AGN where molecular outflows are typically larger than ionized outflows by a factor of $10 - 10^3$. 
    \item The molecular gas reservoirs are depleted and evacuated from the nuclear region on timescales of $10^{6.0} - 10^{7.6}$ years, which agrees with other studies of gas evacuation times. We show that the ionized outflows, rather than the molecular outflows, are the primary drivers of the gas excavation. This level of dominance by the ionized outflows is uncommon in AGN feedback systems.
    \item The gas evacuation occurs on similar timescales as predicted for the AGN duty cycle, which describes the length of time over which the AGN is continuously active. By comparing these timescales, we conclude that AGN outflows are an effective means of clearing the inner few hundred parsecs of cold gas reservoirs in this moderate luminosity Seyfert galaxy.
\end{enumerate}

\begin{acknowledgements}

The authors would like to thank the anonymous referee for the constructive and detailed feedback. J.F. would like to thank Dr. Almudena Alonso-Herrero for graciously sharing her ALMA data on NGC 3227. 

Much of the data presented in this work are based on observations with the NASA/ESA Hubble Space Telescope and were obtained from the Mikulski Archive for Space Telescopes (MAST), which is operated by the Association of Universities for Research in Astronomy, Incorporated, under NASA contract NAS5-26555. These observations are associated with program No. 16246 (\href{https://archive.stsci.edu/proposal\_ search.php?mission=hst\&id=16246}{https://archive.stsci.edu/proposal\_ search.php?mission=hst\&id=16246}).
Support for program No. 16246 was provided through a grant from the STScI under NASA contract NAS5-26555. The specific observations used in this analysis can be accessed via DOI:  \dataset[10.17909/cm86-me24]{\doi{10.17909/cm86-me24}}.

This research has made use of NASA’s Astrophysics Data System. IRAF is distributed by the National Optical Astronomy Observatories, which are operated by the Association of Universities for Research in Astronomy, Inc., under cooperative agreement with the National Science Foundation.
This research has made use of the NASA/IPAC Extragalactic Database (NED), which is operated by the Jet Propulsion Laboratory, California Institute of Technology, under contract with the National Aeronautics and Space Administration.
Some of the observations used in this paper were obtained with the Apache Point Observatory 3.5-meter telescope, which is owned and operated by the Astrophysical Research Consortium.
\end{acknowledgements}

\appendix
\label{sec: appendix}
\restartappendixnumbering
\renewcommand{\thefigure}{A\arabic{figure}}
\setcounter{figure}{0}

\begin{figure*}[t]
\centering  
\includegraphics[width=0.34\linewidth]{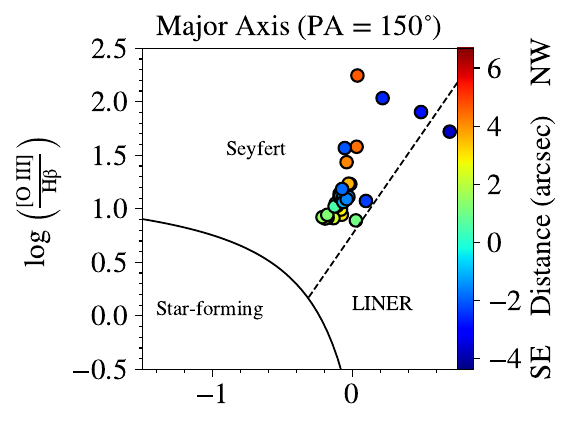}
\hspace{.05mm}
\includegraphics[width=0.31\linewidth]{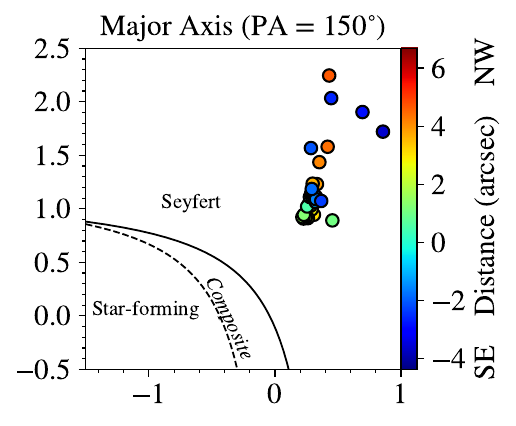}
\hspace{.05mm}
\includegraphics[width=0.31\linewidth]{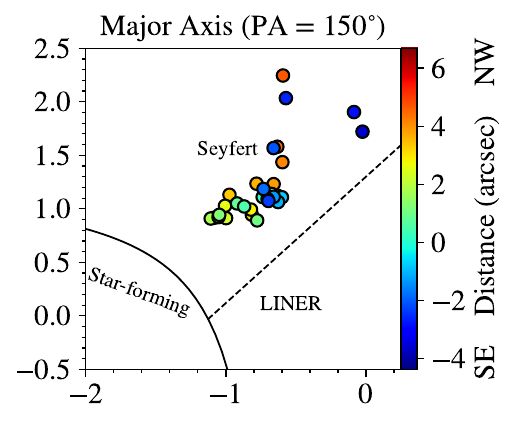}
\includegraphics[width=0.34\linewidth]{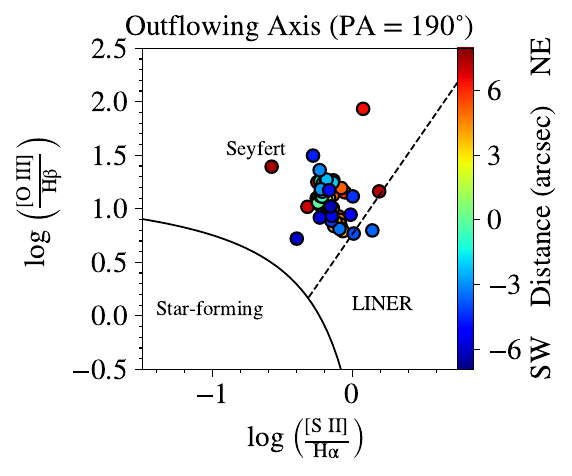}
\hspace{.05mm}
\includegraphics[width=0.31\linewidth]{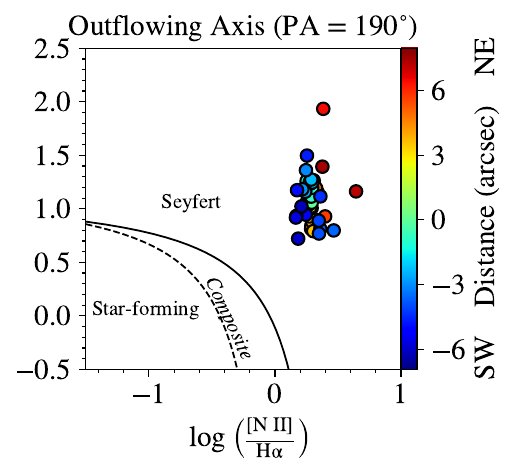}
\hspace{.05mm}
\includegraphics[width=0.31\linewidth]{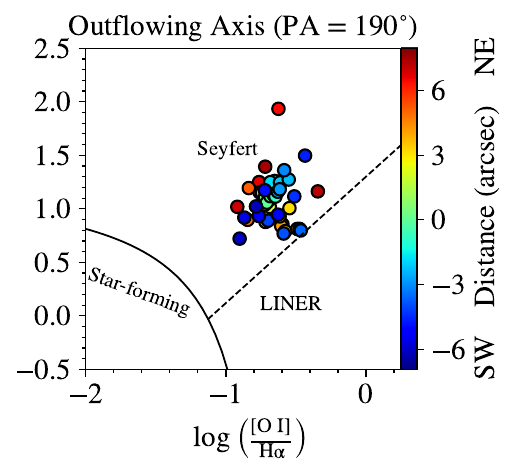}

\caption{BPT ionization diagrams for APO KOSMOS observations along the major (PA = 150\arcdeg, top) and outflowing (PA = 190\arcdeg, bottom) axes. Positive distance refers to the north. The presence of triangle and plus shapes alongside the circles in plots of the major axis indicate second and third components, respectively, in the Gaussian fits.}
\label{fig: other BPTs}

\end{figure*}

In \S \ref{sec: BPT}, we use BPT diagrams to analyze the ionization source along the galactic minor axis of NGC~3227. We use the KOSMOS spectrograph to take data along two other axes of importance: the galactic major axis, and the outflow axis along which the NLR outflows propagate most strongly. The orientation of these other two slits are shown in dotted lines in Figure {fig: OIII contours}. Although the analysis in this paper does not involve these two slits, this section of the Appendix discusses our interpretation of their associated BPT diagrams as shown in Figure \ref{fig: other BPTs}.

\section{Major Axis (PA = 150\texorpdfstring{\arcdeg}{})}
The major axis contains more kinematically complex regions than those observed along the minor axis, and consequently there are a number of three-component fits that are plotted in its BPT diagrams, which is seen in the top row of Figure \ref{fig: other BPTs}. The additional number of components can likely be attributed, at least in part, to the rotational motion which is much more visible along the major axis where it is at its highest projection along our line of sight, compared to the minor axis which suppresses our visibility of rotational motion.

Nevertheless, along this axis, we see clear indication that the gas is ionized by the AGN at virtually all recorded distances, which extend out to $6''$ NW. This may be surprising when comparing to the major axis slit placement in Figure {fig: OIII contours}, which appears to show minimal NLR emission along the major axis beyond $\sim2''$ NW. KOSMOS may have been able to pick up the emission at greater distances than WFC3 because of its larger spatial scale that surveys a greater area to create the spectrum at each slit position. 

\section{Outflows Axis (PA = 190\texorpdfstring{\arcdeg}{})}
The outflows axis is the orientation which most comprehensively envelopes the NLR outflows, as seen in Figure {fig: OIII contours}. Unlike the major and minor axes, the emission from the outflows axis is associated with LINER signatures at distances both NE and SW of the SMBH, according to the leftmost plot in the bottom row of Figure \ref{fig: other BPTs}.  However, the other two plots in the bottom row of Figure \ref{fig: other BPTs} show significantly less overlap with the LINER regime. 

As shown in Figure \ref{fig: 3227}, there is a strong presence of star-forming H~II regions in the immediate vicinity of the SMBH, represented in the figure's inset by red coloring. It is likely that contamination from this emission is creating this effect in the BPT diagram.

\facilities{HST(STIS, WFC3), APO (KOSMOS), Gemini North (NIFS), ALMA}


\software{Cloudy \citep{ferland13}, MultiNest \citep{feroz19}, Astropy \citep{astropy22}}


\bibliography{bibbo}{} 
\bibliographystyle{aasjournal}

\restartappendixnumbering


\end{document}